\documentclass[runningheads]{llncs}
\usepackage{graphicx}
\usepackage{graphics}
\usepackage{tabularx,booktabs}
\usepackage{hyperref}
\usepackage{multicol}
\usepackage{subcaption}
\usepackage[a4paper,left=3cm,right=3cm,top=2.5cm,bottom=2.5cm]{geometry}
\usepackage{siunitx}
\begin{document}
\title{Investigating similarities and differences between South African and Sierra Leonean school outcomes using Machine Learning}

\author{Henry Wandera\inst{1}
\and
Vukosi Marivate \inst{2}
\and
David Sengeh\inst{3}}
\authorrunning{H Wandera et al.}
%
\institute{University of Pretoria \\
\email{u17253129@tuks.co.za}\\
\and
Council for Scientific and Industrial Research  \\
\email{vukosi.marivate@cs.up.ac.za}\\
\and
Government of Sierra Leone\\
\email{dsengeh@statehouse.gov.sl}}
\maketitle              
\begin{abstract}
Available or adequate information to inform decision making for resource allocation in support of school improvement is a critical issue globally. In this paper, we apply machine learning and education data mining techniques on education big data to identify determinants of high schools' performance in two African countries: South Africa and Sierra Leone. The research objective is to build predictors for school performance and extract the importance of different community and school-level features. We deploy interpretable metrics from machine learning approaches such as SHAP values on tree models and odds ratios of LR to extract interactions of factors that can support policy decision making. Determinants of performance vary in these two countries, hence different policy implications and resource allocation recommendations.

\keywords{Education  \and Policy-making \and Machine learning.}
\end{abstract}
\section{Introduction}

We investigate the identification of factors that influence learning outcomes for schools in South Africa (SA) and Sierra Leone (SL). We use diverse data sets including national examination outcomes, important features about schools and communities, as well as national statistics office data to build predictive models that inform education policy. Specially, we investigate the use of interpretable machine learning approaches to extract factors that improve productiveness of models that support public policy decision making. Through this approach, we can identify factors that are important not only for predicting of learning outcomes, but also for illuminating the likely factors that policymakers should consider to improve education outcomes. We compare and contrast our results to factors that have been identified, from literature, applied in other countries using methods such as qualitative surveys. From this, we then make suggestions for education policy.

Africa’s postcolonial education continues to face underlying access and quality challenges including high dropout rates, poor quality of learning and teaching, lack of educational resources, a shortage of trained teachers, poor infrastructure and unavailable teaching and learning materials. These challenges are worse in economically poorer countries. This work seeks to use data analytics to explore important features which can be associated with well performing schools at a national level. The results and analyses can be used to improve resource allocation and decision making by school administrators and education policymakers.

Different groups of researchers with distinct research objectives have conducted related work \cite{baker2009state,romero2010educational} in the field of Educational Data Mining (EDM) by applying statistical, machine learning, and data mining methods to different types of educational data. EDM work has been used to help teachers manage classes, improve students' learning and predict performance. 
Previous research work predominantly applied EDM techniques in the Western world using data sources such as e-learning, learning management systems and intelligent tutoring systems data to education, while Africa lags conspicuously behind. 
This work focuses on Africa and explores the role played by community and school-level factors in determining the performance of high schools in two countries that are geographically, economically and socio-politically different; South Africa and Sierra Leone. The research questions include:
\begin{itemize}
    \item Does the school environment influence performance of students in schools?
    \item Does the community environment influence performance of students in schools?
    \item Which school features are most relevant in determining overall school performance?
    \item What are the characteristics of the best performing schools in each country?
    \item How do school features in Sierra Leone and South Africa differ in determining performance?
\end{itemize}
 We consider the overall performance of schools in the final national senior secondary examinations, but apply Logistic Regression (LR) and tree-based machine learning algorithms to train classifiers and extract significant school features as required in the above-mentioned questions. We use these models because they can potentially predict nonlinear relationships and their results can easily be interpreted by school stakeholders and policymakers.

The rest of this paper is organized as follows. Section \ref{dataset} explains the type of datasets used. Section \ref{methodology} discusses the methodology. Section \ref{results} presents the predictive results from the experiments. Section \ref{interpretation} presents the most significant factors for both countries and discusses their differences and similarities. Section \ref{policy} presents recommendations and the policy implications of the identified factors. Section \ref{literature} and \ref{limitations} discusses prior work and limitations respectively, and Section \ref{conclusion} summarizes this study.

\section{Datasets} \label{dataset}
\subsection{South Africa}
Table ~\ref{tab:sa-variables} shows a summary of the South African dataset. Four main 2016 datasets are used in this study: the school performance dataset obtained from the South African Department of Basic Education, school locations, schools master lists and the results of a 2016 community survey (obtained from Statistics South Africa). The school performance dataset contains final exam results extracted from the 2016 National Senior Certificate (NSC) school performance pdf reports. The 2016 schools master list data for every province was acquired from the Department of Basic Education website. These master lists were merged with school performance files using unique Education Management Information System (EMIS) school codes. The community survey acquired in csv format provides socio-economic information about different households. This is a large-scale survey that targeted approximately 1.3 million households with the objective of providing population and household statistics at municipal level.

\begin{table}[]
\caption{List of SA variables categorized according to their separate datasets which were merged to form one final dataset that is used in the analysis. The final dataset has 32 variables and 5302 secondary schools across South Africa. It contains both qualitative and quantitative data. Qualitative data is good for exploring household open reactions whereas quantitative data from schools such as pass rate, and number of teachers or students, is good for answering quantitative questions and deriving other calculated variables. \textbf{Note:} For all households living in the same local municipality, we calculate the most frequent overall service quality rating in all distinct local municipalities and assign it to all the schools within the same local municipality. This means that if a municipality has the most number of households with no televisions or its water rating is poor, the schools or students in that municipality are assumed not to have access to televisions and have poor water services. This assumption is based on the current school quintile system in SA \cite{dieltiens2014quintile}}.
\label{tab:sa-variables}
\begin{tabular}{@{}lll@{}}
\toprule
\textbf{Variable name}      & \textbf{Data type} & \textbf{Description}                           \\ \midrule
\textbf{School performance dataset} &               &                                                \\
EMIS                        & Integer           & Education Management Information System number \\
Name                        & String        & Name of the school                             \\
Province                    & Category        & Province where the school is located           \\
District                    & Category        & District where the school is located           \\
Quintile                    & Category          & Quintile of the school                         \\
2016 exam \% achieved        & Float        & Pass rate for 2016 ($>$= 50 was pass and  $<$50 was fail)\\
                            &               &                                                \\
\textbf{School master list dataset} &               &                                                \\
EMIS                        & Integer           &  Education Management Information System number\\
Latitiude                   & Float         &  GIS latitude                                  \\
Longititude                 & Float         &  GIS longitude                                 \\
Municipality                & Category        &  Municipality where the school is located      \\
Urban\_Rural                & Category           & Is the school in a rural or urban area         \\
Educators\_2016            & Integer           &  Number of teachers                            \\
Learners\_2016             & Integer           &  Number of Students                            \\
Student\_Teacher\_Ratio      & Float        &  Ratio of students to teachers - calculated variable \\
                            &               &                                                \\
\textbf{2016 Community Survey dataset}   &               &                                                \\
MunicDiff                   & Category           & Difficulties facing the municipality  \\
RateWater                   & Category           & Rating of the overall quality of the water services \\
RateElectricity             & Category           & Rating of the overall quality of the electricity supply services \\
RateToilet                  & Category           & Rating of the overall quality of toilet/sanitation services \\
RateHospital                & Category           & Rating of the overall quality of the local public hospital \\
WaterAccess                 & Category           & Access to safe water supply drinking service \\
Toilet                      & Category           & Main type of toilet facility used    \\
ToiletLocation              & Category           &  The main toilet facility in the dwelling/yard/outside the yard \\
MainDwellType               & Category           & Main dwelling that the household currently lives in \\
SafetyInDay                 & Category           & Safety during the day                 \\
SafetyInDark                & Category           & Safety when it is dark                \\
ElectrInterrupt             & Category           & Interruption in electricity in the past 3 months \\
EnergyLight                 & Category           &  Main source of energy for lighting  \\
HeadHH\_Age\_at\_RefNight   & Integer           & Night Age of household head          \\
HeadHH\_Sex                 & Category          & Sex of household head                 \\
HHgoods\_tv                 & Category           & Household television                 \\
Hhgoods\_radio              & Category           & Household radio                      \\
HHgoods\_dvd                & Category           & Household DVD/Blu-ray player         \\
Internet\_cellphone         & Category           & Internet-Any place via a cellphone . \\ 
\bottomrule
\end{tabular}
\end{table}

\subsection{Sierra Leone}
For Sierra Leone, 3 categories of 2018 datasets were acquired from the Directorate of Science, Technology and Innovation and the Ministry of Basic and Senior Secondary Education (MBSSE): School exam results, 2018 school census data, and Teachers data (a summary of the datasets is provided in Table \ref{tab:sl-variables}). The exam datasets contain the 2018 West African Senior School Certificate Examination (WASSCE) average grades for every school. The 2018 school survey dataset was carried out in all pre-primary, primary, junior and senior secondary schools. It contains school-level features to help the government determine the total number of schools by level, type, location, facilities, furniture and enrolment for the purpose of informing decision makers in the implementation of the Free Quality School Education and to support the Education Sector Plan 2018-20. In this work, we only extract features of senior secondary schools. The teachers dataset is anonymized by excluding their names and other identifying individual information but contains details of their sex, experience, salary source and EMIS code of their schools.

\begin{table}[]
\caption{List of SL variables grouped according to their datasets. After merging these datasets, the final dataset has 37 variables and 162 schools spread across the entire country. We removed over 50 variables from the original dataset because they are either repetitive or highly correlated, and some have many missing values - where more than 50\% of the schools did not provide the information. \textbf{Note:} Each row in the teachers dataset represents a particular individual not the school. We performed counts after grouping teachers by the EMIS code of the schools. This approach provides quantitative variables such as number of teachers per school, number of female or male teachers, and the average teachers service years in every school which were merged with schools}
\label{tab:sl-variables}
\begin{tabular}{@{}lll@{}}
\toprule
\textbf{Variable name}         & \textbf{Data type} & \textbf{Description}                                                                           \\ \midrule
\textbf{2018 School survey}    & \textbf{}          & \textbf{}                                                                                      \\
emiscode                       & integer                & unique school identification provided by MBSSE                                 \\
school\_name                   & string             & name of the school                                                                             \\
latitude                       & float              & school latitude                                                                                \\
longitude                      & float              & school longitude                                                                               \\
sum\_enrol                     & integer                & number of students in the school                                                               \\
remoteness                     & category                & accessibility of the school                                                                    \\
idregion                       & category             & region                                                                                         \\
iddistrict                     & category             & district                                                                                       \\
school\_owner                  & category             & owner of the school                                                                            \\
approval\_status               & category                & has the school been approved by the government                                                 \\
financial\_support             & category                & does the school receive financial support                                                      \\
sh\_fenced                     & category                & is the school fenced                                                                           \\
boarding                       & category                & is it a boarding school                                                                        \\
sch\_garden                    & category                & does the school have a garden                                                                  \\
internet                       & category                & does the school have internet                                                                  \\
drink\_water                   & category                & does the school have drinking water                                                            \\
computers                      & integer                & number of computers in the school                                                               \\
avail\_latrine\_fac            & category                & are toilet facilities available                                                                \\
private\_cubicle               & category                & are there private cubicles                                                                     \\
drink\_water\_source           & category                & what is the source of the drinking water                                                       \\
library                        & category                & does the school have a library                                                                 \\
sci\_lab                       & category                & does the school have science lab                                                               \\
canteen                        & category                & does the school have a canteen                                                                 \\
rec\_facilities                & category                & does the school have a recreation facility                                                     \\
elec\_grid                     & category                & does the school have an electric grid                                                          \\
ssstot                         & integer                & social studies teaching/learning materials                                                     \\
bstot                          & integer                & science teaching/learning materials                                                            \\
mathstot                       & integer                & mathematics teaching/learning materials                                                        \\
basic\_comp\_skills            & category                & do the students have basic computer skills                                                     \\
                               &                    &                                                                                                \\
\textbf{WASSCE exams datasets} & \textbf{}          & \textbf{}                                                                                      \\
emis\_code                     & integer                & unique school identification provided by MBSSE                                 \\
schName                        & string             & name of the school                                                                             \\
papers\_passed                       & float              & percentage of papers passed ($>$= 50 was pass and  $<$50 was fail) \\
                               &                    &                                                                                                \\
\textbf{Teachers dataset}      & \textbf{}          & \textbf{}                                                                                      \\
emis\_code                     & integer                & unique school identification provided by MBSSE                                 \\
teacher\_time                  & category                & does the teacher work part-time or full time                                                   \\
teacher\_prof\_qual            & category                & professional qualification of the teacher                                                      \\
teacher\_aca\_qual             & category                & academic qualification of the teacher                                                          \\
teacher\_service\_years        & integer                & teacher experience - in years.                                                                 \\
teacher\_salary\_source        & category                & who pays the teacher (government, private or community)                                        \\
teacher\_sex                   & category                & gender of the teacher                                                                          \\ \bottomrule
\end{tabular}
\end{table}

\subsection{Differences and similarities in the datasets}
This research focused on the impact of school-level and community-level determinants of school performance and comparing their influence on performances across the two countries. However, the nature of the datasets is different not only in size, but also in coverage and content. The variables in the South African dataset contain community-level factors with only 5 school level features: pass rates, location of the school, quintile, and number of students and teachers. Unlike the former dataset, the Sierra Leone dataset is enriched with school level features and more teachers' details. The distinction of these datasets is derived from the types of surveys that were conducted; one is community based and the other is entirely school based. Since we were using existing datasets, we failed to get similar datasets across these countries because of the differences in policies and interests, and students also sit for different set of exams. Contextually, SA school master lists are similar to SL annual school surveys, but lack the details of more school features and teachers information. The SL dataset also lacks community level details. We investigate which features are important in each country and make recommendations for the collection of more granular education datasets.

\section{Methodology}\label{methodology}
Unlike other machine learning algorithms such as support vector machines and neural networks which have the ability to learn and model non-linear and complex relationships or boundaries but are difficult to interpret\cite{romero2010educational}, we use LR and tree-based machine learning algorithms such as eXtreme Gradient Boosting (XGBoost), Random Forests (RF) and Decision Trees (DT) to model complex relationships in datasets and predict whether a school will pass or fail (school outcome).

The school outcome for Sierra Leone was measured by the percentage of papers students passed in 2018 WASSCE exams, while for South Africa, it was measured by the percentage of students who passed the 2016 Matric exams. In order to develop binary classifiers, two categories were used in both countries: ``fail" and ``pass". For SA, schools with pass rate less than 50\% were labelled as ``fail" and those with pass rate greater or equal to 50\% were labelled as ``pass". For SL, schools with percentage of papers passed less than 50\% were labelled as ``fail" and those with percentage of papers greater or equal to 50\% were labelled as ``pass".

The explanatory variables (x) for the models included community level and school level factors such as security, housing, electricity, student-teacher ratios, computers, latrines and internet (see table \ref{tab:sa-variables} and \ref{tab:sl-variables}). The dependent variable (y) was the school outcome (pass or fail).

\subsection{Models}
\textbf{Decision trees} \cite{quinlan1987simplifying,quinlan1986induction}. This algorithm mimics a human level thinking by representing attributes in a tree-like structure. Each node represents an attribute, each link between nodes represents a decision rule, and the output is represented by the leaf nodes. The best attribute in the dataset is placed at the root of the tree. The training set is then split into subsets. Subsets are made in such a way that each subset contains data with the same value for an attribute. These steps are then repeated for each subset of the training set until leaf nodes are found in all the branches of the tree. However, trees are unstable, prone to overfitting and require taking optimal choices at every node.
\\
\textbf{Random forest} \cite{breiman2001random}. This is a bagging algorithm that uses ensemble learning techniques by training every tree independently and collecting the various decision trees whose results are aggregated into one final result through voting mechanisms. Bagging is a method where different models are trained by resampling the training data to make the resulting model more robust.
Random forests outperform a single decision tree by solving the problem of overfitting and reducing variance. 
\\
\textbf{XGBoost algorithm} \cite{chen2016xgboost}. This algorithm applies bagging and boosting technique. It trains the different models by resampling the data but subsequent models are trained while taking errors made by previously trained models into account. This is called boosting and it reduces the bias whereas bagging reduces variance. This algorithm was selected to address bias and high variance issues faced by tree-based algorithms. Tree models can handle data which is not normally distributed; they form visualisable and interpretable relationships within data and can handle missing data. They apply `if-then' rules which can easily be interpreted by policy makers. End users such as policy makers prefer prediction models that provide hidden actionable insights that are easy to understand, interpret and validate. This helps school administrators and policy makers to provide proactive feedback and suggest resources to particular schools or students. 
\\
\textbf{Logistic regression} \cite{xing2015participation}. This is a regression analysis method for predicting binary dependent variables. It is used to explain the relationship between one dependent binary variable and one or more numeric independent variables. This algorithm was selected because it is a probability/risk estimator, it helps to understand the impact of an independent variable on the dependent variable. LR provides probabilities and odds to measure how likely it is that something will occur. Odds ratios represent the constant effect of a predictor X, on the likelihood that one outcome will occur.
\subsection{Interpretability of ML models}
We use the Predictive, Descriptive, Relevant (PDR)
framework in \cite{murdoch2019interpretable} to guide the modelling, extraction and visualization of significant factors influencing school outcomes.
Although predictive accuracy was used to measure model performance, achieving high accuracy would not be enough because the aim is to extract, investigate and validate the relevance of model results. SHAP (SHapley Additive exPlanations) values \cite{lundberg2017unified} and odds ratios \cite{szumilas2010explaining} were used to interpret models' compositions because of their ability to increase model transparency by indicating how much each predictor contributes. Goodman and Kruskal's gamma \cite{davis1967partial} was used to measure and validate the association between variables, x and y on an ordinal scale, while the Kruskal-Wallis test \cite{kruskal1952use} was used for x numeric variables against y using the P-value of 0.05. To improve the accuracy of the models, a stratified 10-folds cross-validator approach was used to train and evaluate our models on every strata. 
We apply the above-mentioned methods to pinpoint and compare the impacts of the factors in each country to inform unique policy formulation and implications. For instance, whether schools should provide more computers than teachers to improve performance in Sierra Leone? However, estimating education production functions requires modelling interactions of more than one factor because a combination of various factors comes into play to influence the performance of students in schools.

\section{Predictive Results} \label{results}
This section unveils the predictive power of our built models and the most significant determinants for each country. Table \ref{tab:model-performance} reports the average performance of the models (in 10 experiments) on the test cases. Experiments were conducted after tuning the hyperparameters of every model. XGBoost outperformed other classifiers with the accuracy of 65.5\% and 75.5\% , and specificity of 78.5\% and 82.9\% for SL and SA respectively. Specificity means the ability of the model to correctly identify good schools with pass rates or percentage of papers passed greater or equal to 50\%. For sensitivity, the XGBoost model was ranked third in correctly identifying SL failing schools with 67.7\% performance after RF with 71.1\% and LR which led with 73.3\%.

The best two models which were selected for feature extraction were considered based on 2 criteria: the best Area Under Curve (AUC) achieved for tree models, and a non tree-based based model. We used AUC to measure the separability because it can show how much a model is capable of distinguishing between classes. The higher the AUC, the better the model. XGBoost had the best overall AUC of 77.3\% and 82.1\% for SL and SA respectively. The LR model which was second with AUC of 79.5\% was selected because of its relatively comparable performance and for purposes of extracting the odds ratios and validating XGBoost top features on a linear scale. It should be noted that the training dataset was balanced, which is why we also relied on accuracy as one of the top performance metrics. It is also easily understood by non-technical stakeholders.

In general, the performance of the models across countries varied because they were each trained in different contexts, with different size of datasets and number of variables. Considering the AUC metric, the XGBoost algorithm emerged best in all countries, while LR emerged third after DT and RF in SL and SA respectively. XGBoost is an optimized distributed gradient boosting algorithm which can have better performance than RF and DT if parameters are correctly tuned \cite{nielsen2016tree}.

\begin{table}[]
\centering
\caption{Average performance of models (in \%) on South Africa and Sierra Leone datasets.}
\label{tab:model-performance}
\begin{tabular}{@{}lllll|lllll@{}}
\toprule
& \multicolumn{4}{c|}{\textbf{South Africa}}
& \multicolumn{4}{c}{\textbf{Sierra Leone}}\\
\cmidrule(lr){2-5}\cmidrule(lr){6-10}
\textbf{Model} & \multicolumn{1}{c}{\textbf{Accuracy}}
& \multicolumn{1}{c}{\textbf{Sentivity}}
& \multicolumn{1}{c}{\textbf{Specificity}}
& \multicolumn{1}{c|@{}}{\textbf{AUC}}
&
& \multicolumn{1}{c}{\textbf{Accuracy}}
& \multicolumn{1}{c}{\textbf{Sentivity}}
& \multicolumn{1}{c}{\textbf{Specificity}}
& \multicolumn{1}{c@{}}{\textbf{AUC}}\\

\midrule
XGBoost             & 75.5     & 67.7      & 82.9        & 82.1 &  & 65.5     & 63.2      & 78.5        & 77.3 \\
Logistic regression & 72.3     & 73.3      & 71.4        & 79.5 &  & 61.2     & 48.1      & 77.2        & 60.3 \\
Decision trees      & 75.1     & 59        & 90.6        & 79.1 &  & 61.2     & 59.3      & 63.6        & 63.1 \\
Random forests      & 70.5     & 71.1      & 69.9        & 78.8 &  & 55.1     & 44.4      & 68.2        & 55.7 \\ \bottomrule
\end{tabular}
\end{table}

\section{Interpretation} \label{interpretation}
\begin{figure}[]
\caption{\label{fig:sa-featureimportance}}
  \begin{subfigure}{\columnwidth}
    \centering\includegraphics[width=\columnwidth]{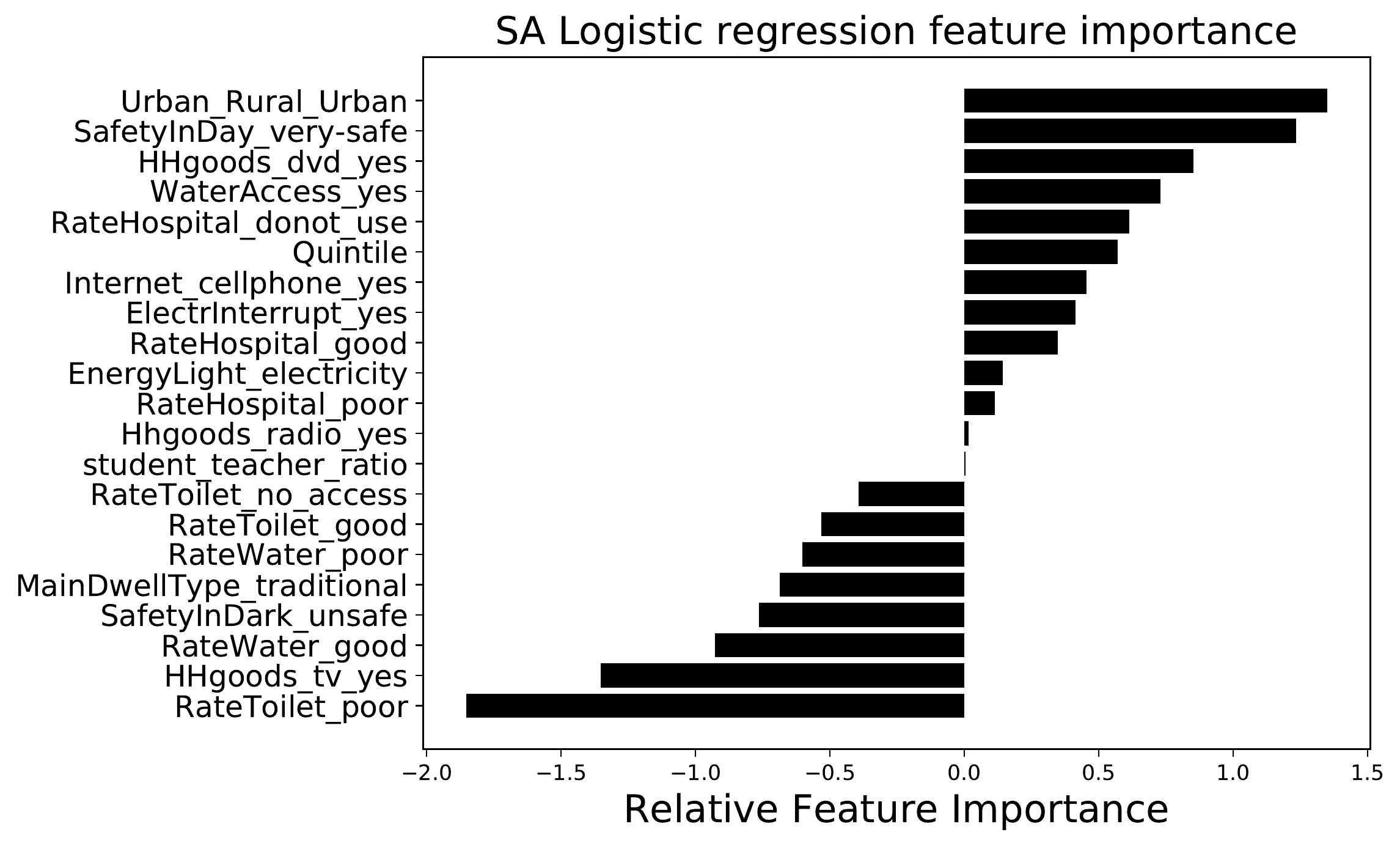}
    \caption{A bar chart showing the impact of every feature in the logistic regression model. Urban schools, very safe safety in the day, access to water, quintiles and other features whose bars are on the right increased the odds of passing while poor toilets, traditional dwellings and insecurity at night reduced the odds of passing.}
  \end{subfigure}
  \begin{subfigure}{\columnwidth}
    \centering\includegraphics[width=\columnwidth]{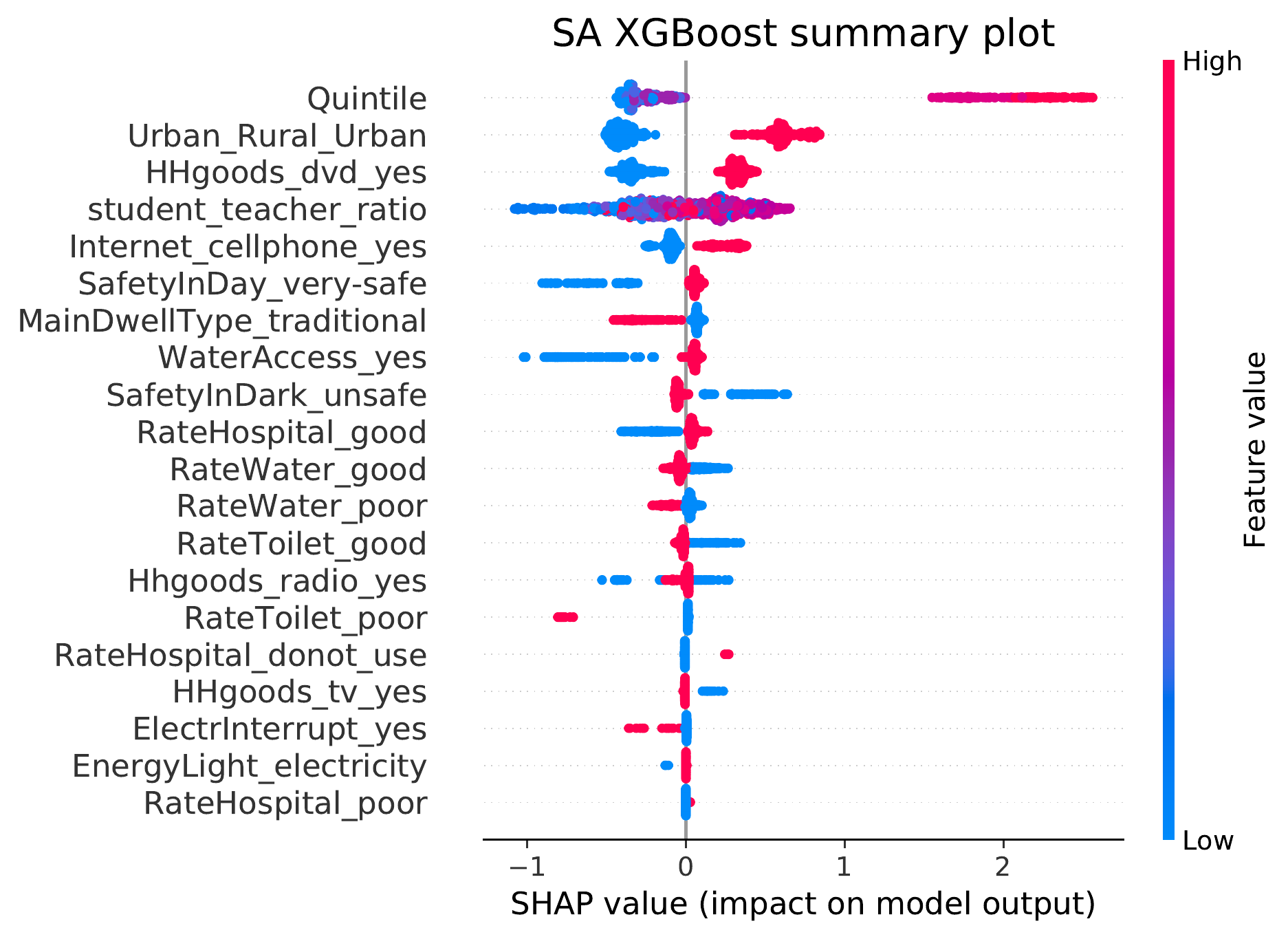}
    \caption{A summary plot showing the distribution of every feature impact in the XGBoost model. The color represents the feature value (red represents high and blue represents low). Low quintiles lowered performance in schools, high cases of cellphone internet usage increased performance and high cases of traditional dwellings were associated to decrease in the performance of students in schools.}
  \end{subfigure}
\end{figure}

Figures \ref{fig:sa-featureimportance} and \ref{fig:sl-featureimportance} show the features that were found to be most relevant in determining the school outcomes in both countries; however, the feature importance rankings varied for each model. For Logistic regression, large positive values of relative feature importance implies higher importance of a particular feature in predicting whether students will pass in the school, while negative values imply higher importance in predicting the failure rate. 
\subsection{South Africa}
School quintiles, location (rural or urban), student-teacher ratio, good hospitals, very safe security during the day, access to water and availability of DVDs, and cellphone Internet were found to be significant in predicting good performance in South Africa. Poor toilets, electricity interruptions, and traditional dwellings, had higher importance in predicting the fail outcome in South Africa. In order to validate these determinants, we investigated their distributions in schools which scored 100\% (404 strong schools) and in schools with less than or equal to 40\% pass rates (565 struggling schools). Table \ref{tab:sa-cat-prevalence} shows the number of schools and their percentage representation in each category. We split struggling schools into two groups namely; weak schools (with 0-20\% pass rate) and fair schools (with 21-40\% pass rate). There were 119 weak schools and 446 fair schools raising a percentage of 2.2\% and 8.4\% of the whole dataset respectively, while 7.6\% were strong schools.

Quintiles were ranked number 1 by XGBoost and number 6 by LR in having a high positive impact on performance. School quintiles range from 5 to 1 with 5 at the top in having a better socioeconomic status and 1 as the worst. Results show that 64.1\% of the strong schools belong to quintile 5, only 2 fair schools were quintile 5, and there was no weak quintile 5 school. 7.4\% of strong schools belonged to the quintile 1 category compared to the 42.8\% of fair schools and 56.3\% of weak schools. Researchers in \cite{ogbonnaya2019quintile} also found similar performance disparities.

Good hospitals were also found significant with its unit increase contributing 41.5\% change to the odds of pass vs fail. 91.1\% of strong schools were located in communities with good hospital rating compared to 77.6\% of fair and 66.4\% of weak schools.

Results show that schools located in areas mostly with traditional dwellings were negatively affected in achieving high pass rate outcomes. Only 2.7\% of strong schools were located in these areas compared to 28.5\% and 37.8\% of fair and weak schools. This feature was found to reduce the odds of pass vs fail by -49.7\%.

Availability of televisions and radios was not significant in explaining the variations in school outcomes. It was found that more than 88.0\% of all schools (strong, fair and weak) were in communities where households accessed or used these services.

\subsection{Sierra Leone}
\begin{figure}[]
\caption{\label{fig:sl-featureimportance}}
  \begin{subfigure}{\columnwidth}
    \centering\includegraphics[width=\columnwidth]{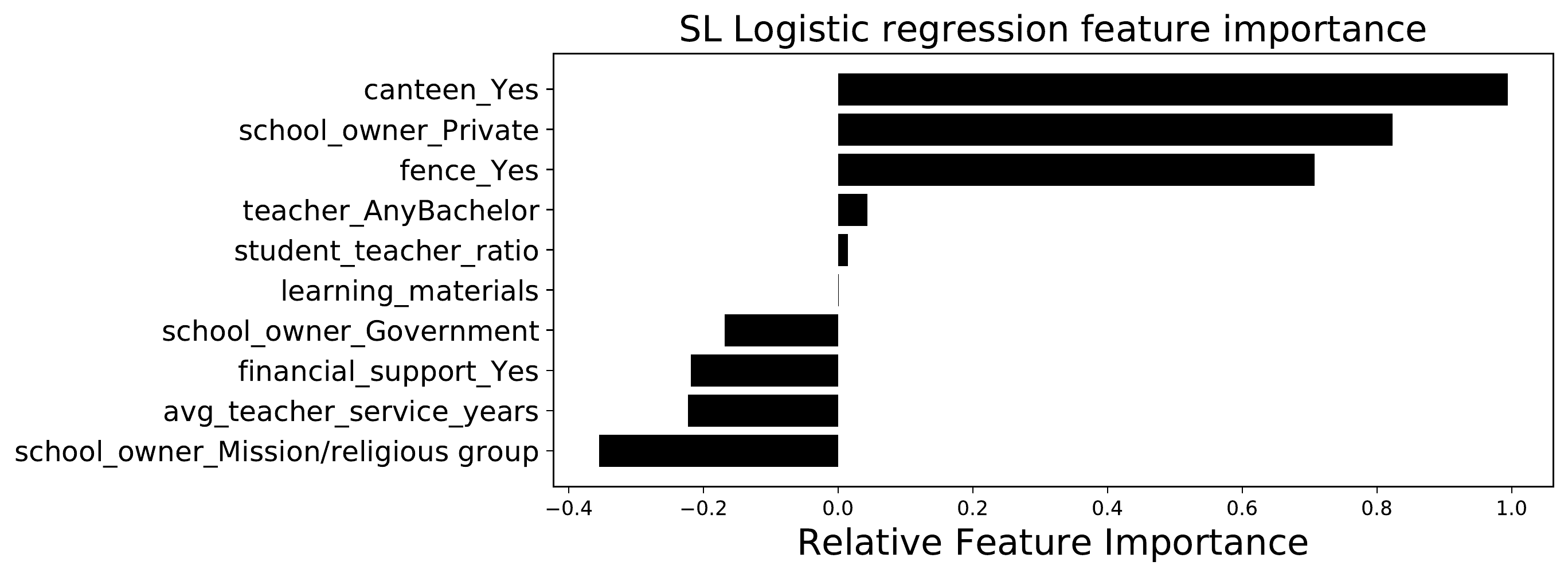}
    \caption{This bar chart was extracted from the LR model and it represents Sierra Leone feature rankings. Availability of canteen, school fence and private school status increased the odds of passing in schools while government or mission schools with more financial support reduced the odds of passing.}
  \end{subfigure}
  \begin{subfigure}{\columnwidth}
    \centering\includegraphics[width=\columnwidth]{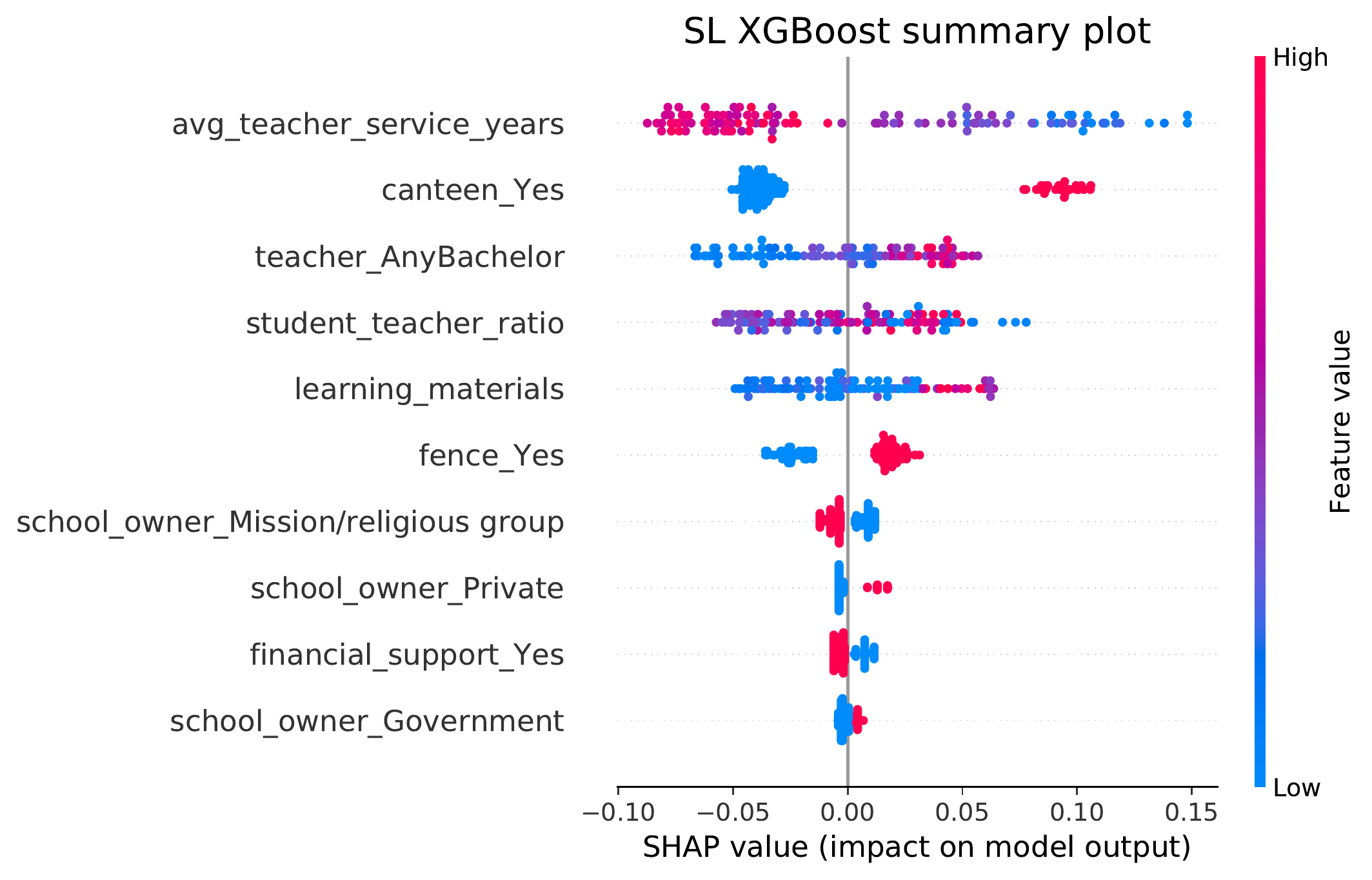}
    \caption{A summary plot showing the distribution of Sierra Leone features as extracted from the XGBoost model. High values are represented by the red colour and low values are in blue. High values of average teaching experience were associated to low performance and schools with no fences were associated with low students' performance.}
  \end{subfigure}
\end{figure}

Results show that private school ownership, favourable location (region), availability of canteen, fence and electricity, and the number of teachers with any bachelor degrees were the most significant features in predicting the pass outcome. Mission and government school ownership, financial support were found with a higher importance in predicting the fail outcome (see Figure \ref{fig:sl-featureimportance}).

\begin{figure}[]
\caption{South Africa and Sierra Leone pruned decision trees. \label{fig:trees}}
  \begin{subfigure}{\columnwidth}
    \centering\includegraphics[width=\columnwidth]{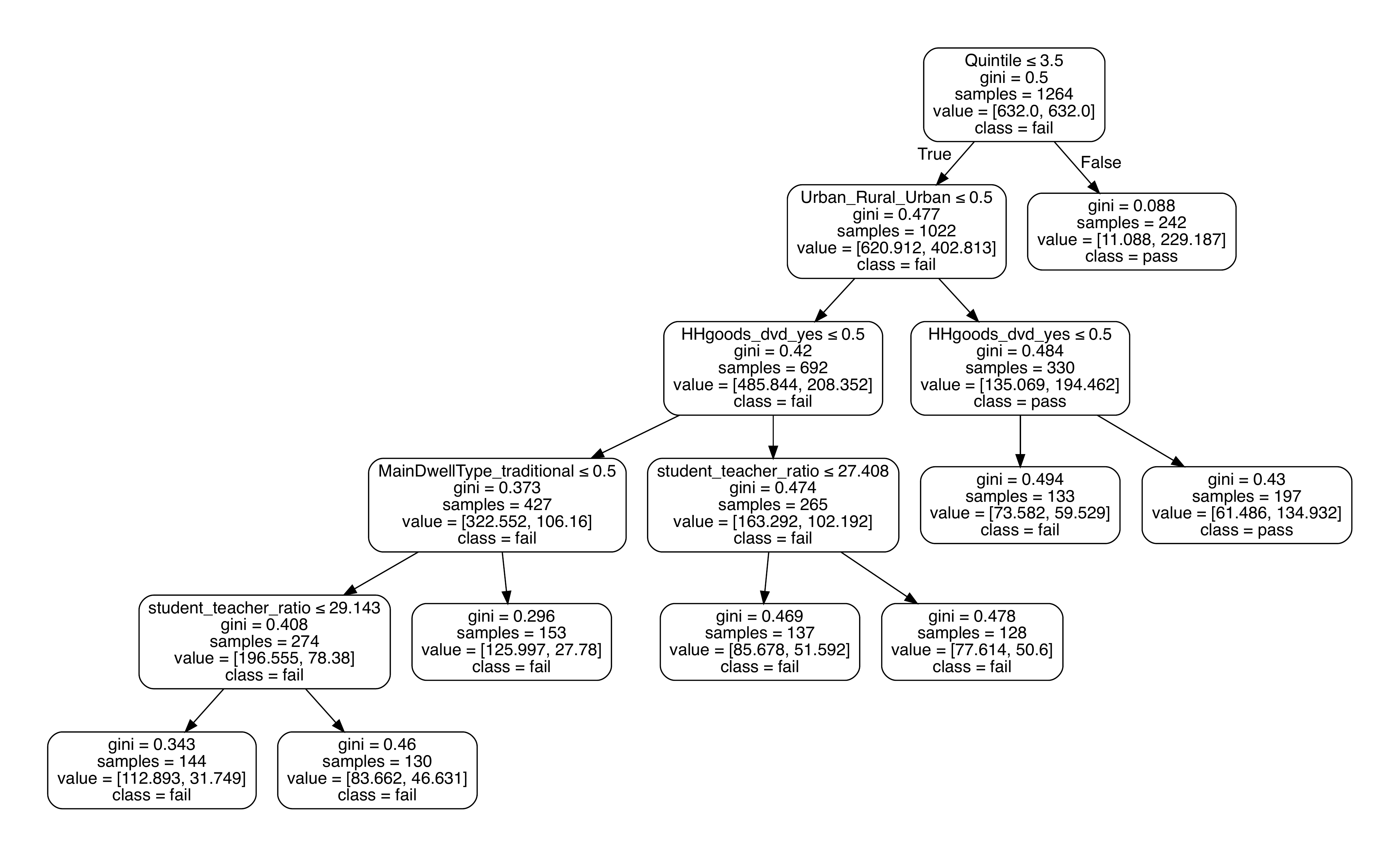}
    \caption{A pruned decision tree for South Africa: The split was formed on quintiles followed by the rural-urban school location divides. Schools with quintile 4 and 5 were categorised in the pass category with a Gini impurity of 0.088 whereas urban schools located in areas where most households had DVDs were also categorised in the pass category.}
    \label{sa-decision-tree}
  \end{subfigure}
  \begin{subfigure}{\columnwidth}
    \centering\includegraphics[width=\columnwidth]{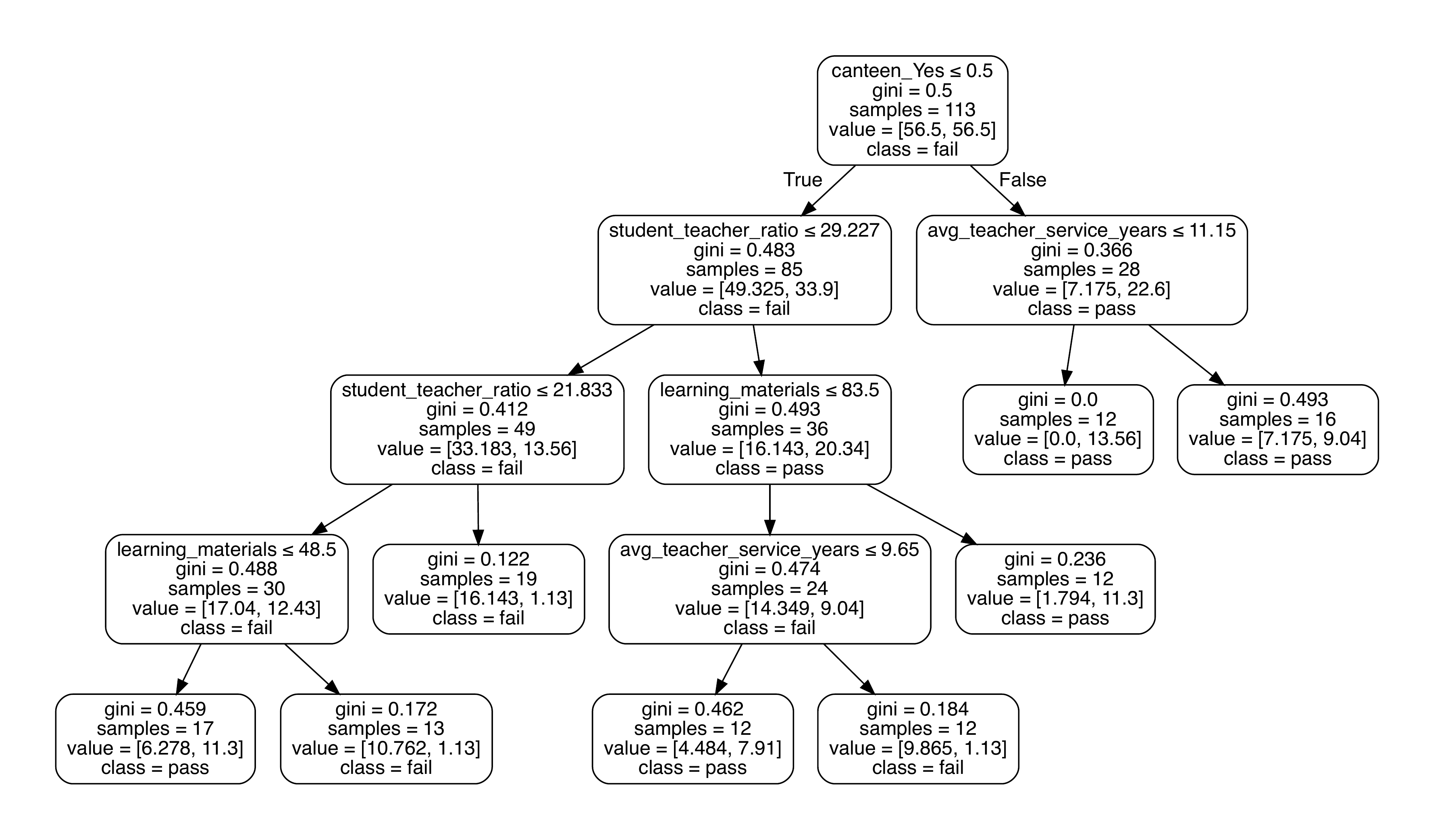}
    \caption{Decision tree for Sierra Leone: Like XGBoost and LR models, the availability of canteens in Sierra Leone schools was also ranked in the first position by the decision tree model. Schools with canteens where the average teachers experience was less than 11.15 years (at Gini impurity of 0.366), were considered in the pass category with 0 Gini impurity.}
    \label{sl-decision-tree}
  \end{subfigure}
\end{figure}

Availability of canteens in schools was ranked in the first and second position (by LR and XGBoost respectively) in having a positive impact on the school outcomes. This is also evident in the SL decision tree (see Figure \ref{sl-decision-tree}). A  canteen is a restaurant provided by a school for its students or staff to get food either freely or at a cost. This shows that it is paramount to provide feeding as a social safety net in addition to its other benefits like reducing school drop-outs discussed in \cite{ibrahim2017role}. Results showed that 74.4\% of schools which had canteen passed while only 38.3\% of schools with no canteens passed. However, the prevalence of canteens was mostly in the Western region of the country where 50.9\% of the schools had these facilities compared to other regions with less than 0.1\% of their schools.

Most of the variables in the SL dataset failed to explain the variations in the school outcomes. Table \ref{tab:sl-category-summary} and Figure \ref{fig:sl-numeric-summary} shows the frequencies and distributions of categorical and numeric variables respectively. Results show that most of the schools regardless of their outcome had almost the same resources. For categorical features, the association of the following variables with the response variable was not significant with their Goodman-Kruskal's gamma values at P-value $>$ 0.05: remoteness, mixed school, boarding, development plan, drinking water,drinking water source, library, approval status, shift status, garden, internet, private cubicle, science lab, recreational facilities, generator and basic computer skills. At 0.05 significance level, results of ANOVA and Kruskal-Wallis tests show that the following numeric variables had identical populations for both pass and fail school outcomes thus were not good predictors of school outcomes: total number of latrines, computers, counsellors, chalkboards, textbooks and student-teacher ratio. This shows that there are other underlying challenges within the school education system for policy considerations in addition to the provision of basic amenities. 

\subsection{Comparison of South Africa and Sierra Leone results}
Determinants of school outcomes varied in both countries because of the differences in their datasets compositions. The SA dataset was primarily community based as opposed to the SL dataset which covered more school characteristics. For South Africa, results showed the influence of economic factors on school outcomes. For instance, schools located in communities with formal dwellings, good hospitals, access to water, electricity and ICT tools performed better than those in disadvantaged areas. For Sierra Leone, private schools and schools with canteens, and a fence performed better than schools without these amenities. 

Results of the models show that schools in urban areas had a high positive impact on performance. In South Africa, this feature was ranked first and second for LR and XGBoost respectively. Analytically, 85.9\% of the strong schools were in urban areas compared to 12.6\% of weak schools and 22.0\% of fair schools. This feature was significant with Goodman-Kruskal's gamma ($\gamma$) association coefficient of 0.74 at P-value $<$ \num{2.2e-16} in explaining the variations in school outcomes. Similarly, Sierra Leone schools in the Western region performed better than those in other regions because this area comprises the national capital Freetown and its surrounding is considered urban where most schools had better amenities. It was discovered that 83.6\% of Western schools had electricity unlike other regions with less than 45.9\% electricity coverage. Energy poverty limits the use of teaching resources and classroom materials such as printers for making assignment copies, computers and access to internet for research purposes \cite{ouf2018effect}. Results in South Africa also showed that higher values of electricity interruptions had a negative impact on the outcomes.

Researchers in \cite{alokan2013rural} and \cite{gibbs2000challenge} also found that the performance of rural schools is not outstanding not simply because of their rural locations but because of factors like low teachers' salaries, qualifications and experience. Moreover, on average, schools in the Western Urban Region had more qualified teachers. For example, in Western Urban in SL, 13 teachers with BEduc and 22 teachers with any bachelor degrees compared to schools in other regions with 8 and 10 teachers respectively. Similarly, the Southern and North Western regions were the worst performers with 38.6\% and 43.1\% of papers passed. They were found to have the least number of qualified teachers with degrees and lowest electricity connectivity. The North Western region was also found to have the least total number of learning materials (textbooks).

Some ICT tools were found to be determinants of school outcomes. For instance, in Sierra Leone, the number of computers in good schools in the western region was 3 times more than that in other regions. In South Africa, availability of DVDs in households was ranked 3rd by both XGBoost and LR. Results show that 84.9\% of strong schools were located in communities where most households have DVDs, unlike the 33.4\% of fair schools and 16.8\% of weak schools. The $\gamma$ value of Access to/use of cellphone internet was 0.63 (P-value $<$ \num{2.2e-16}) indicating quite a strong association with the school outcome.
The LR model also provided a 57.5\% change in odds of pass vs fail for every unit increase in access to cellphone internet. Work in \cite{gaikwad2010interactive,gibson2004visions,sharif2006association} also associates these services to better performance and improved access to information which in this case maybe relevant to school candidates.

Results also showed that security in communities and schools plays an important role in explaining school outcomes as low cases of very-safe safety during the day could have a negative impact on performance. In South Africa, the LR model results (in Table \ref{tab:sa-odd-ratios}) indicated that very-safe safety during day was associated to increasing the odds of passing over failing by 3.43 (243.7\%), while communities with unsafe nights were linked to reducing the odds of passing by 53.36\%, when other factors remain constant. Similarly, in Sierra Leone, results showed that availability of school fences was associated with more cases of passing in schools as 55\% of fenced schools passed compared to the 38\% unfenced schools which managed to pass. A school fence established to prevent or control access or exits can be considered as a security measure with the aim of not only protecting the students, staff, and resources but also reducing events of students escaping from schools and student misbehavior \cite{servoss2017school}.
\begin{figure}[t]
\begin{multicols}{2}
    \includegraphics[width=.5\textwidth]{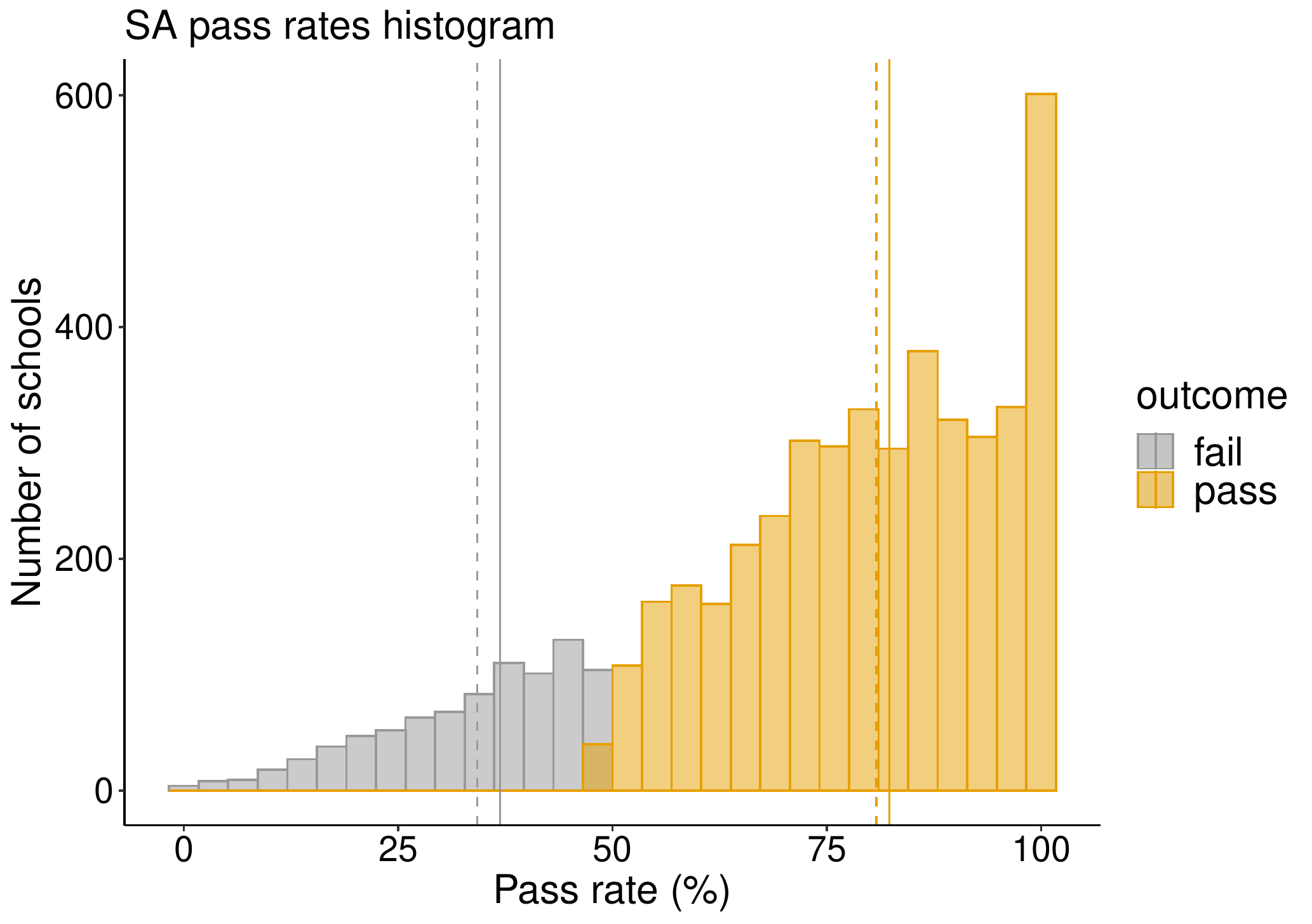}\par 
    \includegraphics[width=.5\textwidth]{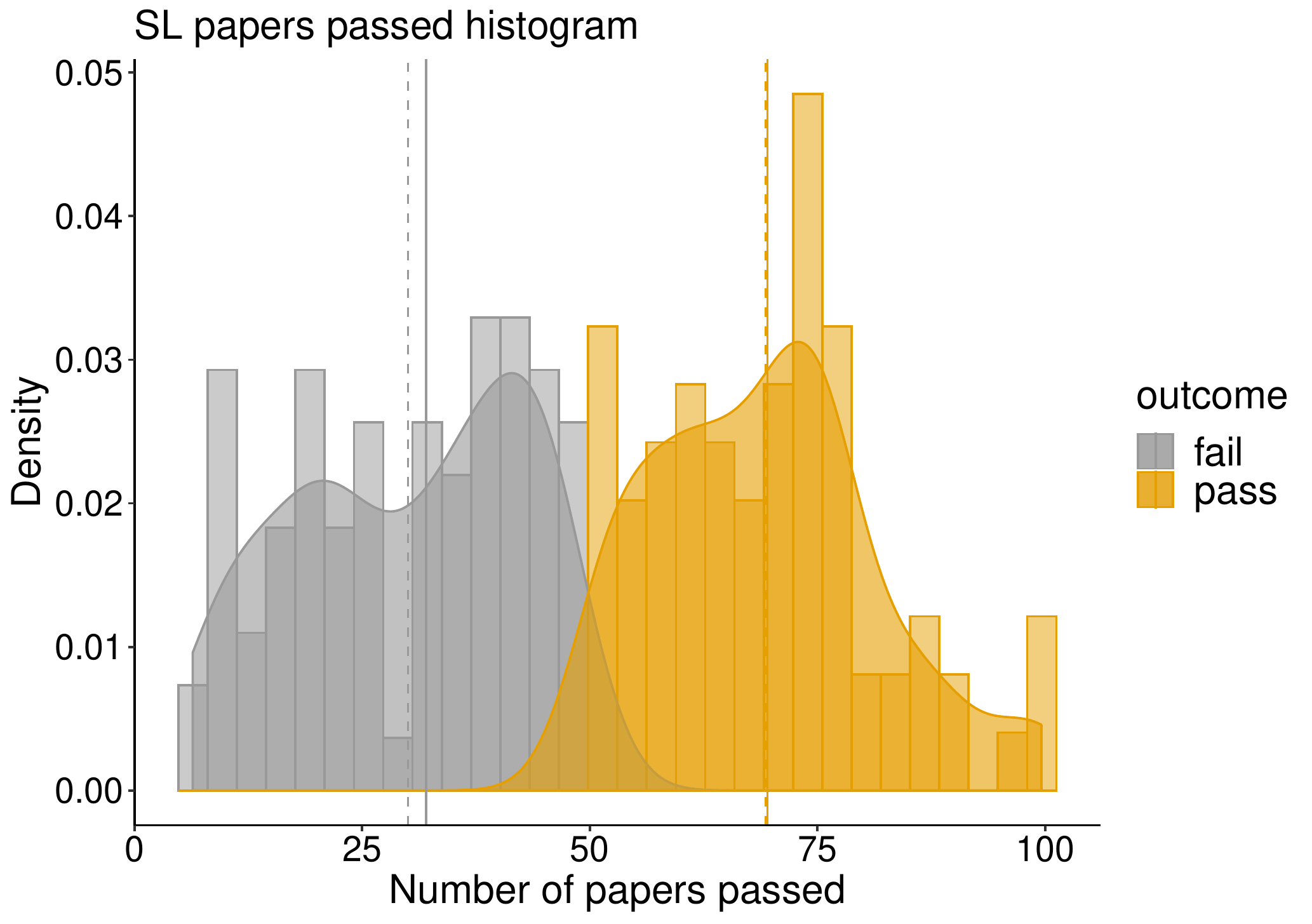}\par 
\end{multicols}
\begin{multicols}{2}
    \includegraphics[width=.5\textwidth]{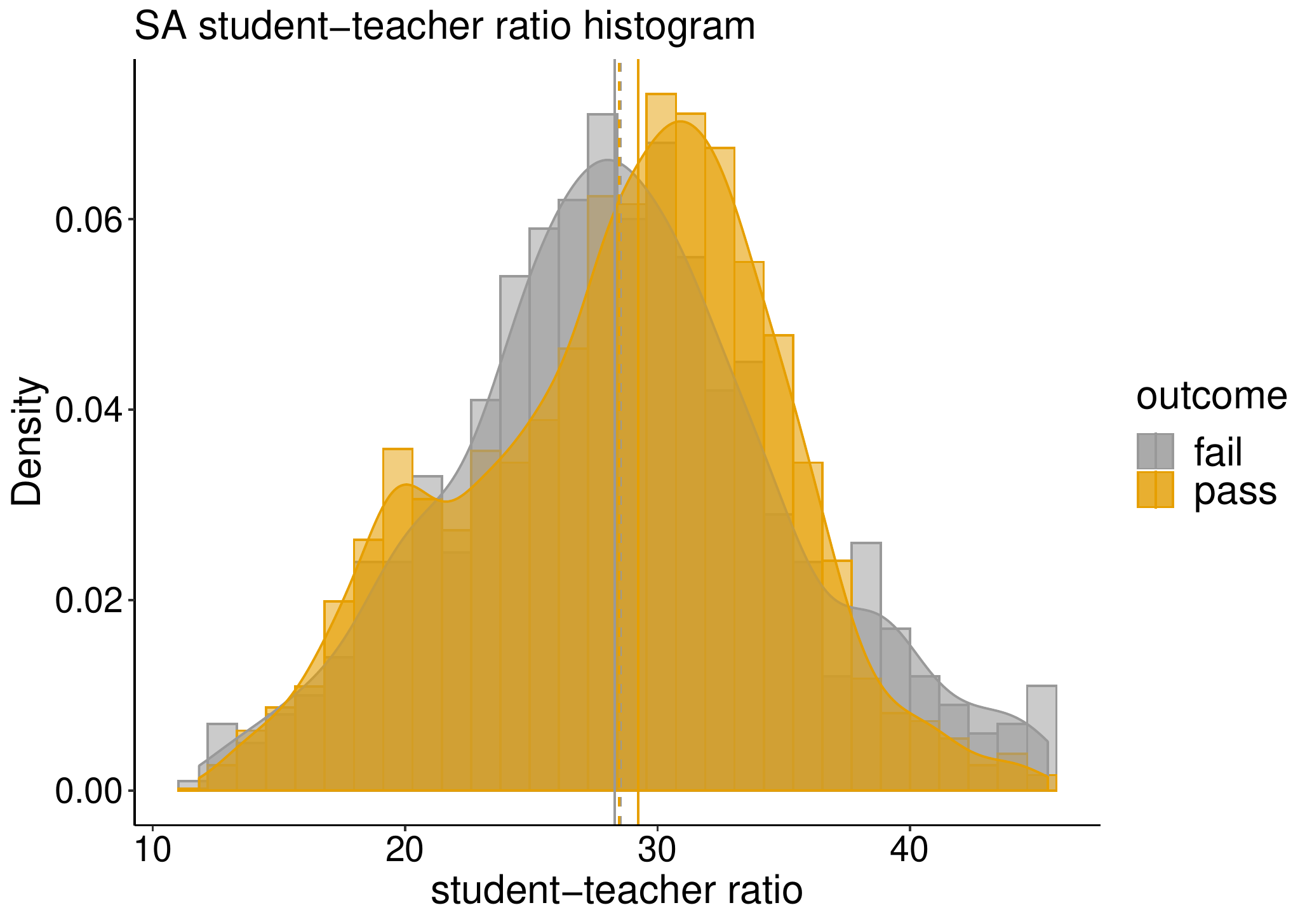}\par
    \includegraphics[width=.5\textwidth]{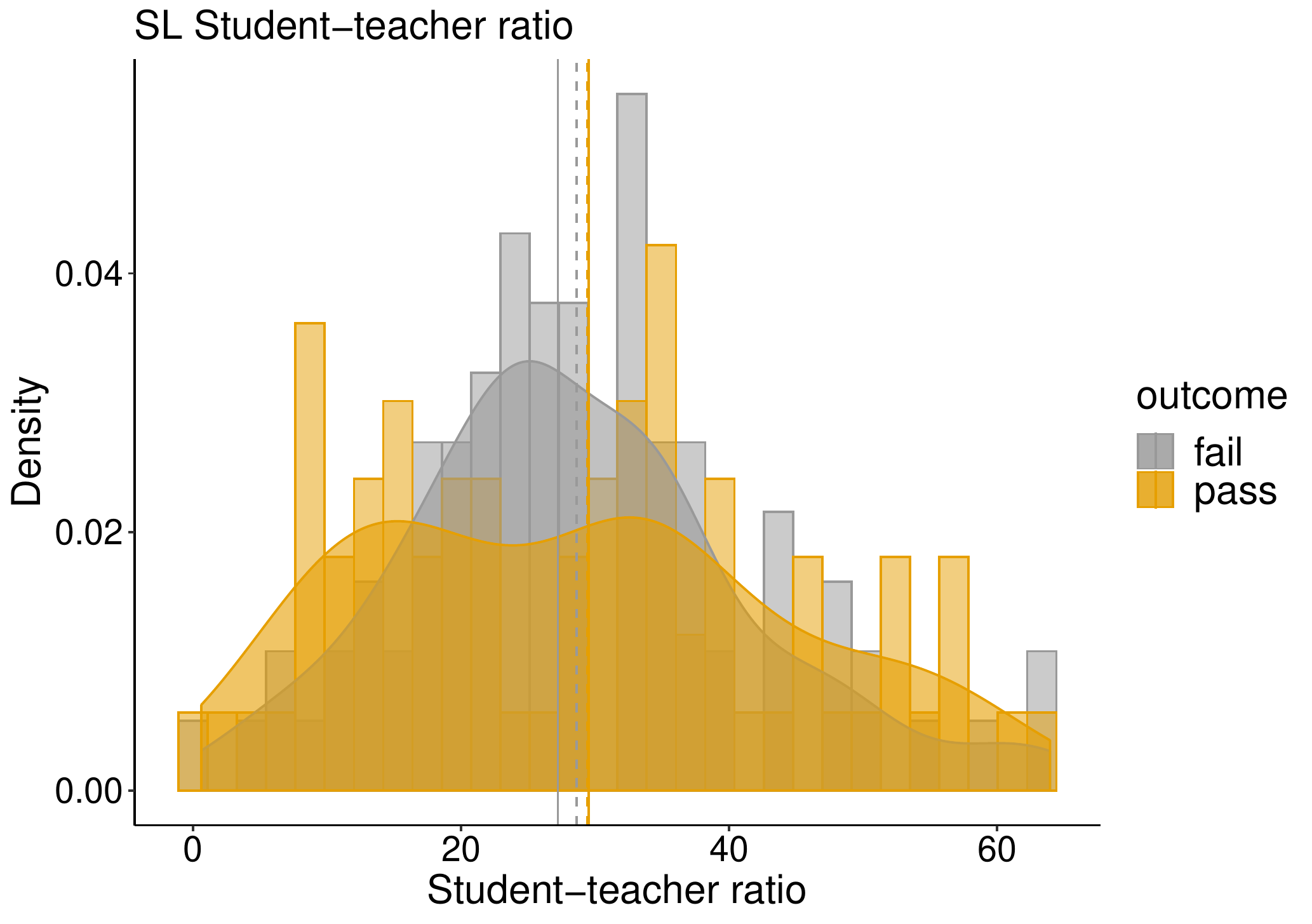}\par
\end{multicols}
\caption{Histogram showing school outcomes, number of teachers and students for SA on the left and SL on the right. Dashed lines indicate the mean in each category while the solid line indicate the median values. For SA, the average pass rate in schools which failed (in grey) was 34.2 (median = 36.9) compared to 80.7 (median = 82.3) in schools which passed. For SL, the average number of papers passed for schools in the fail category was 30\% (median = 32) while schools which passed attained 69.3\% (median = 69.5). There was no significant difference in the student-teacher ratio between the good and struggling schools in all countries. The mean = 28 in all South African schools and 29 in Sierra Leone schools.}
\label{fig:sa-numeric-summary}
\end{figure}

\section{Policy Implications}\label{policy}
This section reflects on the relationships between resources devoted to schools and student performance. Such understanding informs decision making for appropriate and effective education policies that seek to improve learning outcomes. Results show that there is no strong or consistent relationships between variations in school resources and students performance. This not only complicates policy-making processes, but also makes it hard to derive an objective method for determining what is enough to improve performance in schools at scale.

For instance, government schools in Sierra Leone employed more experienced and qualified teachers than other schools. Although higher experience of teachers is associated with better school outcomes \cite{adeyemi2008teachers}, this was not the case across the board in government schools. There are many possible reasons for this observed effect including inefficiencies in operations. Increasing resources or incentives in schools with poor leadership and management may not be adequate for improving school outcomes. Alternatively, resource allocation and administration based on well-defined structures tailored to encourage better performance poses a greater likelihood of success. However, there is no clear definition of what adequate support means particularly as it is perceived as a political or economic issue changing with political views and with demands of the national economy\cite{hanushek1997assessing}. While there is no strong relationship between available resources and school performance in government schools, increased resources must be enhanced by other measures including effective school management to see any noticeable impacts.  

A specific example is that schools that were receiving financial support from government were found to be struggling as only 42.7\% passed while 63.2\% of non-supported schools passed. Private schools performed better than government schools which were receiving financial support. How finances are spent is more important than how much is spent. Private schools had canteens, electricity grids, libraries, science labs and were fenced. Their total number of computers was twice the number of computers in other schools. To set policies appropriately, policy-makers would need to understand the circumstances under which these schools operate and how they use government subsidies and resources. Understanding how resources are utilised through specific policies and practices is necessary for achieving the desired teaching and learning outcomes. More canteens, computers and other materials is not sufficient for better performance in schools if they are not matched to address specific individual learning needs, and if there is no capacity built to effectively manage those resources. Schools can be tasked to enforce programmes and processes which guarantee expertise in the management of resources by encouraging financial audits, monitoring and supervision. While the analyses are for all schools within the country, district-level and chiefdom-level analyses are necessary for developing interventions that are tailored to schools for their specific desired outcomes. 

For South Africa, a positive is that the data for performance (using Matric results) are available. It could be made readily accessible by opting for having one of the public outputs of the yearly reporting being machine readable. There is also a need to explore other performance metrics (being careful not to rely on any single number). Examples of data that could be made available per school are: Trends in International Mathematics and Science Study (TIMSS)\footnote{\url{https://timssandpirls.bc.edu/}}, Annual National Assessments (ANA)\footnote{\url{https://www.education.gov.za/Curriculum/AnnualNationalAssessment.aspx}} and the National Benchmark Test (NBT)\footnote{\url{https://www.nbt.ac.za/}} used for some higher education entrance requirements. We do note that the South African NBT is written by students who have to pay to take the test and as such will be skewed. It may still be able to give insights.

Information on resources and outcomes can be made publicly available to everyone through reports with transparent details of how resources are used with provided evidence of their impact on learning outcomes. This new paradigm of using big data research to support education policy-making was previously limited due to lack of enough and interlinked educational datasets. New government policies and processes which promote collecting of various layers of information and data about what is happening in school systems and the local environments will transform the field. The data sources must include socioeconomic background of students and teachers, school facilities, school access and attendance, school management and governance, images of school infrastructure and geographic data, teachers, and data generated by interactions with learning management systems. Regular or real-time data collected from school entities can be used for, supporting student learning, teachers and teaching, and school administration and management. These data, together with its complete and detailed metadata, can also support research in education anonymized and aggregated for the public.

\section{Prior work} \label{literature}
In this section, we discuss 2 education research methods: statistical analysis of factors and qualitative methods which have been predominantly used in most educational research.
Qualitative researchers in \cite{romero2013predicting,blazar2017teacher,al2017students,uddin2014impact,tapia2004instrument} observed students participation in focus groups, depth interviews and group discussions. Their work has an orientation to the social context - they studied subjectivity in students actions, what they think, and how they attach meanings to different educational events. These methods identify causes and provide us with knowledge about how different types of learning take place and propose personalized solutions to improve students learning. However, because there is no way to analyze qualitative data mathematically, it is difficult to investigate causality, quantify solutions, and verify opinions as students have control over what they provide within those contexts.
In the information age, the desire to explore increasing educational datasets generated in schools facilitated the use of various statistical analysis methods. However, these methods sometimes require qualitative methods to explore some findings further. Research in \cite{yildirim2017effects,xiaogao2017research,junco2015predicting,romero2013web,chen2014mining,dekker2009predicting,nebot2006identification} used statistical approaches to formulate insights and discover patterns in data from surveys, simulations and learning tools. Some of the methods used such as analysis of variance are prone to outliers, Linear Regression assume linear relationships within datasets and Neural Networks can be difficult to be interpreted by users or policy-makers.
This research work focused on interpretability by building models that are easy to use, interpret and also be validated by policy-makers. In order to deduce more robust research outputs, we used tree-based models which applies an if-then analysis by mimicking a human approach to decision making and also assumes no existence linear relationships among variables. SHAP values were used to extract patterns and understand different resource variations with performance in order to inform policy-making. Unlike studies which used data about students, school-level and community-level features were used to investigate their associations with the performance in schools since positive learning environments were found to improve student performance in \cite{kraft2016school,steinberg2011student,turner2009influence,schneider1990model}.
\section{Limitations}\label{limitations}
The proposed tree-based algorithms used require correct parameter tuning as models can overfit especially if wrong parameters  are  used \cite{wang2010handling}, for instance, a large  number  of leaves  may  cause  overfitting  and  parameters  are  dependent on  each  other.
Another challenge faced  was  limited  education  data  sources. There was enough exams results data for South African schools from 2016 to 2019 but the dataset mostly contained 2016 community-household features with not enough updated school-level features such as information about school facilities, teachers data and learning materials to explain the educational phenomena. Nevertheless, the performance of schools was potentially high especially in communities where most households had access to basic resources, hence some community features had strong association with the performance in schools. For Sierra Leone, data was strictly from school surveys with school-level details and teachers information about experience, age, qualifications and salary. However, there were many missing values hence the most critical information was not provided to explain most resource variations with the performance in schools. The WASSCE examination data was also not diverse as most schools scored below the average grade. However, the percentage of papers passed was considered to be the measure of school performance.

\section{Conclusion} \label{conclusion}
Large-scale educational data offers unprecedented opportunities for teachers and policy-makers to understand how students learn from different educational events and what takes place in school systems. In this paper we have applied machine learning techniques to learn patterns in existing education datasets purposely to guide resource allocation, inform decision making and educational policy. Like prior statistical analysis findings, our results confirm their relationships, but also new nonlinear interpretable relationships which support that improving the school and community environment can potentially influence the performance of students in schools. However, building capacities to effectively manage resources within these environments is critical to ensure that they particularly serve individual needs required to improve performance. This study should be extended to other African countries to customise educational policies based on the complexities and differences in resource variations with performance in their schools. The results in this paper should be viewed alongside other social scientists and educationists work which explain social issues in education. For instance, how quintile factors, rural-urban divides, and safety issues affect performance. Lastly, more useful data-driven exploratory tools powered by machine learning interpretable models should be developed to assist in viewing these relationships, quantifying resource inputs and school outcomes in a predictive approach for implementation with policy makers who seek to transform education outcomes at a national and local level.
\bibliographystyle{splncs04}
\bibliography{references}
\appendix
\section{Appendix}

\begin{table}[]
\centering
\caption{Descriptive statistics of SL numeric variables such as total number of computers or students with disabilities in schools. It shows both overall means and medians of these variables and in schools that passed and failed. Although there is a difference in the means, median values are identical in all groups due to outliers hence these schools mostly had identical number of resources.}
\label{tab:sl-numeric-summary}
\begin{tabular}{@{}lll|lll|lll@{}}
\toprule
& \multicolumn{2}{c|}{\textbf{overall}}
&& \multicolumn{2}{c|}{\textbf{mean}}
&& \multicolumn{2}{c}{\textbf{median}}\\
\cmidrule(lr){2-3}\cmidrule(lr){5-6}\cmidrule(lr){8-9}
& \multicolumn{1}{c}{\textbf{mean}}
& \multicolumn{1}{c|}{\textbf{median}}
&& \multicolumn{1}{c}{\textbf{fail}}
& \multicolumn{1}{c|@{}}{\textbf{pass}}
&& \multicolumn{1}{c}{\textbf{fail}}
& \multicolumn{1}{c@{}}{\textbf{pass}}\\
\textbf{Variable} \\ \midrule
computers                                         & 6.1   & 0      &  & 4.7         & 7.6         &  & 0             & 0             \\
disabilities                                      & 3.8   & 1      &  & 5.1         & 2.5         &  & 1             & 1             \\
cousellors                                        & 0.9   & 1      &  & 0.8         & 1.1         &  & 1             & 1             \\
total number of latrines                          & 9     & 8      &  & 9.2         & 8.8         &  & 8             & 6             \\
chalkboards                                       & 12.9  & 12     &  & 12.6        & 13.2        &  & 12            & 12            \\
total number of textbooks                         & 92.1  & 36     &  & 75.4        & 110.4       &  & 36            & 32            \\
total number of teachers                          & 28.3  & 24     &  & 27.6        & 28.9        &  & 24            & 25            \\
total number of full-time teachers                & 26.6  & 23     &  & 26.1        & 27.2        &  & 23            & 23            \\
total numbre of part-time teachers                & 1.6   & 0      &  & 1.5         & 1.7         &  & 0             & 0             \\
total number of teachers with any bachelor degree & 15.7  & 13     &  & 14.3        & 17.2        &  & 12            & 16            \\
total number of teachers with BEduc degree        & 13.5  & 10     &  & 13.3        & 13.7        &  & 9             & 11            \\
total number of government teachers               & 14    & 0      &  & 18          & 15          &  & 18            & 15            \\
total number of private teachers                  & 0     & 0      &  & 1.2         & 6.5         &  & 0             & 0             \\
average service years                             & 10.8  & 11.3   &  & 11.2        & 10.4        &  & 11            & 10            \\
papers passed                                     & 48.7  & 47.4   &  & 30          & 69.3        &  & 32            & 69.5          \\
total numbre of students                          & 836.4 & 683    &  & 817.2       & 857.6       &  & 670           & 710           \\
student-teacher ratio                             & 28.2  & 29     &  & 28.6        & 29.4        &  & 27            & 29\\ \bottomrule       
\end{tabular}
\end{table}

\begin{table}[]
\caption{Shows number of schools in 4 groups categorized by quantiles of papers passed and their responses. In the first and second quantile (1Q and 2Q), all schools failed - the ranges of papers passed (\%) were [6.4, 31.3] and (31.3, 47.4] respectively. Unlike 4Q with all schools which passed (top schools), some schools in 3Q failed while others passed. There were 41 schools in 1Q: 39 accessible, 2 in rough terrains and no school in island, and only 3 were boarding schools.}
\label{tab:sl-category-summary}
\begin{tabular}{l|l|l|l|l|l}
\toprule
\multicolumn{1}{c}{} &
\multicolumn{1}{c}{\textbf{1Q}} & \multicolumn{1}{c}{\textbf{2Q}} &
\multicolumn{2}{c}{\textbf{3Q}} & 
\multicolumn{1}{c}{\textbf{4Q}} \\
\cline{2-6}
        & \textbf{{[}6.4, 31.3{]}} & \textbf{(31.3,  47.4{]}} & \textbf{(47.4,  68.6{]}} & \textbf{(47.4,  68.6{]}} & \textbf{(68.6, 99.6{]}} \\
\textbf{Variable}       & \textbf{fail} & \textbf{fail} & \textbf{fail} & \textbf{pass} & \textbf{pass} \\ 
\midrule
remoteness: accessible                & 39                 & 40                 & 3                  & 32   & 37                \\
remoteness: island                    & 0                  & 0                  & 0                  & 2    & 0                 \\
remoteness: rough terrains            & 2                  & 0                  & 1                  & 2    & 4                 \\
school owner: community               & 4                  & 5                  & 0                  & 3    & 1                 \\
school owner: government              & 9                  & 9                  & 0                  & 8    & 8                 \\
school owner: mission group           & 26                 & 24                 & 4                  & 22   & 17                \\
school owner: private                 & 2                  & 2                  & 0                  & 3    & 15                \\
mixed school: boys only               & 1                  & 3                  & 1                  & 4    & 5                 \\
mixed school: girls only              & 1                  & 7                  & 0                  & 2    & 2                 \\
mixed school: mixed                   & 39                 & 30                 & 3                  & 30   & 34                \\
boarding: yes                         & 3                  & 5                  & 1                  & 7    & 5                 \\
boarding: no                          & 38                 & 35                 & 3                  & 29   & 36                \\
development plan: yes                 & 35                 & 31                 & 3                  & 31   & 37                \\
development plan: no                  & 6                  & 9                  & 1                  & 5    & 4                 \\
drinking water: yes                   & 40                 & 34                 & 4                  & 34   & 35                \\
drinking water: no                    & 1                  & 6                  & 0                  & 2    & 6                 \\
drinking water source: borehole       & 17                 & 21                 & 1                  & 14   & 18                \\
drinking water source: hand dug well  & 15                 & 12                 & 1                  & 17   & 8                 \\
drinking water source: pipe-borne     & 9                  & 6                  & 2                  & 5    & 15                \\
drinking water source: stream         & 0                  & 1                  & 0                  & 0    & 0                 \\
library: yes                          & 28                 & 21                 & 3                  & 24   & 29                \\
library: no                           & 13                 & 19                 & 1                  & 12   & 12                \\
canteen: yes                          & 5                  & 4                  & 1                  & 10   & 19                \\
canteen: no                           & 36                 & 36                 & 3                  & 26   & 22                \\
electricity grid: yes                 & 20                 & 22                 & 1                  & 18   & 34                \\
electricity grid: no                  & 21                 & 18                 & 3                  & 18   & 7                 \\
approval status: applied for approval & 0                  & 2                  & 0                  & 1    & 1                 \\
approval status: approved             & 40                 & 35                 & 4                  & 32   & 40                \\
approval status: not approved         & 1                  & 3                  & 0                  & 3    & 0                 \\
shift status: double shift, afternoon & 13                 & 9                  & 0                  & 9    & 12                \\
shift status: double shift, morning   & 0                  & 0                  & 0                  & 1    & 1                 \\
shift status: single                  & 28                 & 31                 & 4                  & 26   & 28                \\
fence: yes                            & 18                 & 21                 & 1                  & 17   & 32                \\
fence: no                             & 23                 & 19                 & 3                  & 19   & 9                 \\
garden: yes                           & 20                 & 28                 & 3                  & 19   & 20                \\
garden: no                            & 21                 & 12                 & 1                  & 17   & 21                \\
internet: yes                         & 4                  & 4                  & 0                  & 5    & 7                 \\
internet: no                          & 5                  & 9                  & 2                  & 4    & 13                \\
internet: unknown                     & 32                 & 27                 & 2                  & 27   & 21                \\
available latrine: yes                & 41                 & 40                 & 4                  & 33   & 40                \\
available latrine: no                 & 0                  & 0                  & 0                  & 3    & 1                 \\
public cubicle: yes                   & 4                  & 6                  & 0                  & 4    & 9                 \\
public cubicle: no                    & 37                 & 34                 & 4                  & 32   & 32                \\
science lab: yes                      & 20                 & 26                 & 3                  & 21   & 24                \\
science lab: no                       & 21                 & 14                 & 1                  & 15   & 17                \\
recreation facility: yes              & 35                 & 33                 & 3                  & 30   & 30                \\
recreation facility: no               & 6                  & 7                  & 1                  & 6    & 11                \\
generator: yes                        & 20                 & 15                 & 2                  & 16   & 23                \\
generator: no                         & 21                 & 25                 & 2                  & 20   & 18                \\
basic computer skills: yes            & 8                  & 12                 & 2                  & 8    & 16                \\
basic computer skills: no             & 33                 & 28                 & 2                  & 28   & 25                \\
financial support: yes                & 35                 & 33                 & 3                  & 29   & 24                \\
financial support: no                 & 6                  & 7                  & 1                  & 7    & 17                \\
region: eastern                       & 5                  & 6                  & 1                  & 9    & 4                 \\
region: north western                 & 11                 & 7                  & 0                  & 9    & 2                 \\
region: northern                      & 5                  & 10                 & 0                  & 2    & 5                 \\
region: southern                      & 13                 & 9                  & 3                  & 6    & 0                 \\
region: western                       & 7                  & 8                  & 0                  & 10   & 30                \\ \bottomrule
\end{tabular}
\end{table}

\begin{table}[]
\centering
\caption{SA Logistic regression odd ratios. For a school in an urban area, the odds of pass vs. fail were by a factor of 3.86 or would increase by 285.74\%}
\begin{tabular}{lrrr}
\toprule
\textbf{Variables} &   \textbf{weights} &  \textbf{odd-ratio} & \textbf{\%Change} \\
\midrule
            RateWater: good & -0.93 &   0.40 &  -60.42 \\
            RateWater: poor & -0.60 &   0.55 &  -45.30 \\
           RateToilet: good & -0.53 &   0.59 &  -41.23 \\
      RateToilet: no-access & -0.39 &   0.67 &  -32.53 \\
           RateToilet: poor & -1.85 &   0.16 &  -84.32 \\
         Urban\_Rural: urban &  1.35 &   3.86 &  285.74 \\
    RateHospital: donot-use &  0.61 &   1.85 &   84.63 \\
         RateHospital: good &  0.35 &   1.42 &   41.51 \\
         RateHospital: poor &  0.11 &   1.12 &   12.12 \\
           WaterAccess: yes &  0.73 &   2.08 &  107.53 \\
 MainDwellType: traditional & -0.69 &   0.50 &  -49.66 \\
     SafetyInDay: very-safe &  1.23 &   3.44 &  243.75 \\
       SafetyInDark: unsafe & -0.76 &   0.47 &  -53.36 \\
       ElectrInterrupt: yes &  0.41 &   1.51 &   51.29 \\
   EnergyLight: electricity &  0.14 &   1.15 &   15.49 \\
            HHgoods\_tv: yes & -1.35 &   0.26 &  -74.14 \\
         HHgoods\_radio: yes &  0.02 &   1.02 &    1.66 \\
           HHgoods\_dvd: yes &  0.85 &   2.35 &  134.61 \\
    Internet\_cellphone: yes &  0.45 &   1.57 &   57.45 \\
                  Quintile &  0.57 &   1.77 &   76.86 \\
     student-teacher ratio &  0.004 &   1.004 &    0.415 \\
\bottomrule
\end{tabular}
\label{tab:sa-odd-ratios}
\end{table}

\begin{table}[]
\centering
\caption{Frequency of categorical features in South African communities with the strong 404 schools which scored 100\%, and 565 struggling schools with $<$=40\% pass rate separated into groups: 119 weak schools (with 0-20\% pass rate) and 446 fair schools (with 21-40\% pass rate). There was no strong school located in a community with no electricity, while 1 weak school (0.8\% of 199) and 4 fair schools (0.9\% of 446) were located in areas with no electricity.}
\label{tab:sa-cat-prevalence}
\begin{tabular}{@{}lll|ll|ll@{}}
\toprule
& \multicolumn{2}{c|}{\textbf{Strong}}
& \multicolumn{2}{c|}{\textbf{Weak}}
& \multicolumn{2}{c}{\textbf{Fair}}\\
\cmidrule(lr){2-3}\cmidrule(lr){4-5}\cmidrule(lr){6-7}
\textbf{Variable}& \multicolumn{1}{c}{\textbf{number}}
& \multicolumn{1}{c|}{\textbf{percentage}}
& \multicolumn{1}{c}{\textbf{number}}
& \multicolumn{1}{c|@{}}{\textbf{percentage}}
& \multicolumn{1}{c}{\textbf{number}}
& \multicolumn{1}{c@{}}{\textbf{percentage}}\\ \midrule
RateElectricity: no-access & 0            & 0.0              & 1    & 0.8       & 4     & 0.9        \\
RateToilet: no-access      & 0            & 0.0              & 0    & 0.0       & 2     & 0.4        \\
EnergyLight: candles        & 0            & 0.0              & 1    & 0.8       & 8     & 1.8        \\
RateToilet: poor           & 1            & 0.2              & 5    & 4.2       & 13    & 2.9        \\
ElectrInterrupt: yes       & 3            & 0.7              & 3    & 2.5       & 10    & 2.2        \\
Hhgoods\_tv: no            & 3            & 0.7              & 4    & 3.4       & 12    & 2.7        \\
WaterAccess: no            & 3            & 0.7              & 21   & 17.6      & 68    & 15.2       \\
RateHospital: donot-use    & 4            & 1.0              & 0    & 0.0       & 3     & 0.7        \\
Hhgoods\_radio: no         & 5            & 1.2              & 14   & 11.8      & 40    & 9.0        \\
RateElectricity: average   & 5            & 1.2              & 4    & 3.4       & 9     & 2.0        \\
RateToilet: average        & 8            & 2.0              & 17   & 14.3      & 46    & 10.3       \\
MainDwellType: traditional & 11           & 2.7              & 45   & 37.8      & 127   & 28.5       \\
RateHospital: poor         & 12           & 3.0              & 0    & 0.0       & 1     & 0.2        \\
RateHospital: average      & 20           & 5.0              & 40   & 33.6      & 96    & 21.5       \\
Quitile: 3                 & 22           & 5.4              & 22   & 18.5      & 104   & 23.3       \\
RateWater: poor            & 24           & 5.9              & 48   & 40.3      & 146   & 32.7       \\
Quitile: 1                 & 30           & 7.4              & 67   & 56.3      & 191   & 42.8       \\
RateWater: average         & 31           & 7.7              & 22   & 18.5      & 34    & 7.6        \\
Quitile: 2                 & 36           & 8.9              & 29   & 24.4      & 146   & 32.7       \\
Quitile: 4                 & 57           & 14.1             & 1    & 0.8       & 3     & 0.7        \\
Urban\_Rural: Rural        & 57           & 14.1             & 104  & 87.4      & 348   & 78.0       \\
HHgoods\_dvd: no           & 61           & 15.1             & 99   & 83.2      & 297   & 66.6       \\
SafetyInDark: safe         & 64           & 15.8             & 27   & 22.7      & 64    & 14.3       \\
SafeInDay: fairly-safe     & 99           & 24.5             & 10   & 8.4       & 38    & 8.5        \\
Internet\_cellphone: yes   & 189          & 46.8             & 10   & 8.4       & 65    & 14.6       \\
Internet\_cellphone: no    & 215          & 53.2             & 109  & 91.6      & 381   & 85.4       \\
Quitile: 5                 & 259          & 64.1             & 0    & 0.0       & 2     & 0.4        \\
SafeInDay: very-safe       & 305          & 75.5             & 109  & 91.6      & 408   & 91.5       \\
SafetyInDark: unsafe       & 340          & 84.2             & 92   & 77.3      & 382   & 85.7       \\
HHgoods\_dvd: yes          & 343          & 84.9             & 20   & 16.8      & 149   & 33.4       \\
Urban\_Rural: Urban        & 347          & 85.9             & 15   & 12.6      & 98    & 22.0       \\
RateWater: good            & 349          & 86.4             & 49   & 41.2      & 266   & 59.6       \\
RateHospital: good         & 368          & 91.1             & 79   & 66.4      & 346   & 77.6       \\
MainDwellType: formal      & 393          & 97.3             & 74   & 62.2      & 319   & 71.5       \\
RateToilet: good           & 395          & 97.8             & 97   & 81.5      & 385   & 86.3       \\
HHgoods\_radio: yes        & 399          & 98.8             & 105  & 88.2      & 406   & 91.0       \\
RateElectricity: good      & 399          & 98.8             & 114  & 95.8      & 433   & 97.1       \\
WaterAccess: yes           & 401          & 99.3             & 98   & 82.4      & 378   & 84.8       \\
ElectrInterrupt: no        & 401          & 99.3             & 116  & 97.5      & 436   & 97.8       \\
HHgoods\_tv: yes           & 401          & 99.3             & 115  & 96.6      & 434   & 97.3       \\
EnergyLight: electricity    & 404          & 100.0            & 118  & 99.2      & 438   & 98.2       \\ \bottomrule
\end{tabular}
\end{table}

\begin{figure}[]
\begin{multicols}{2}
    \includegraphics[width=.43\textwidth]{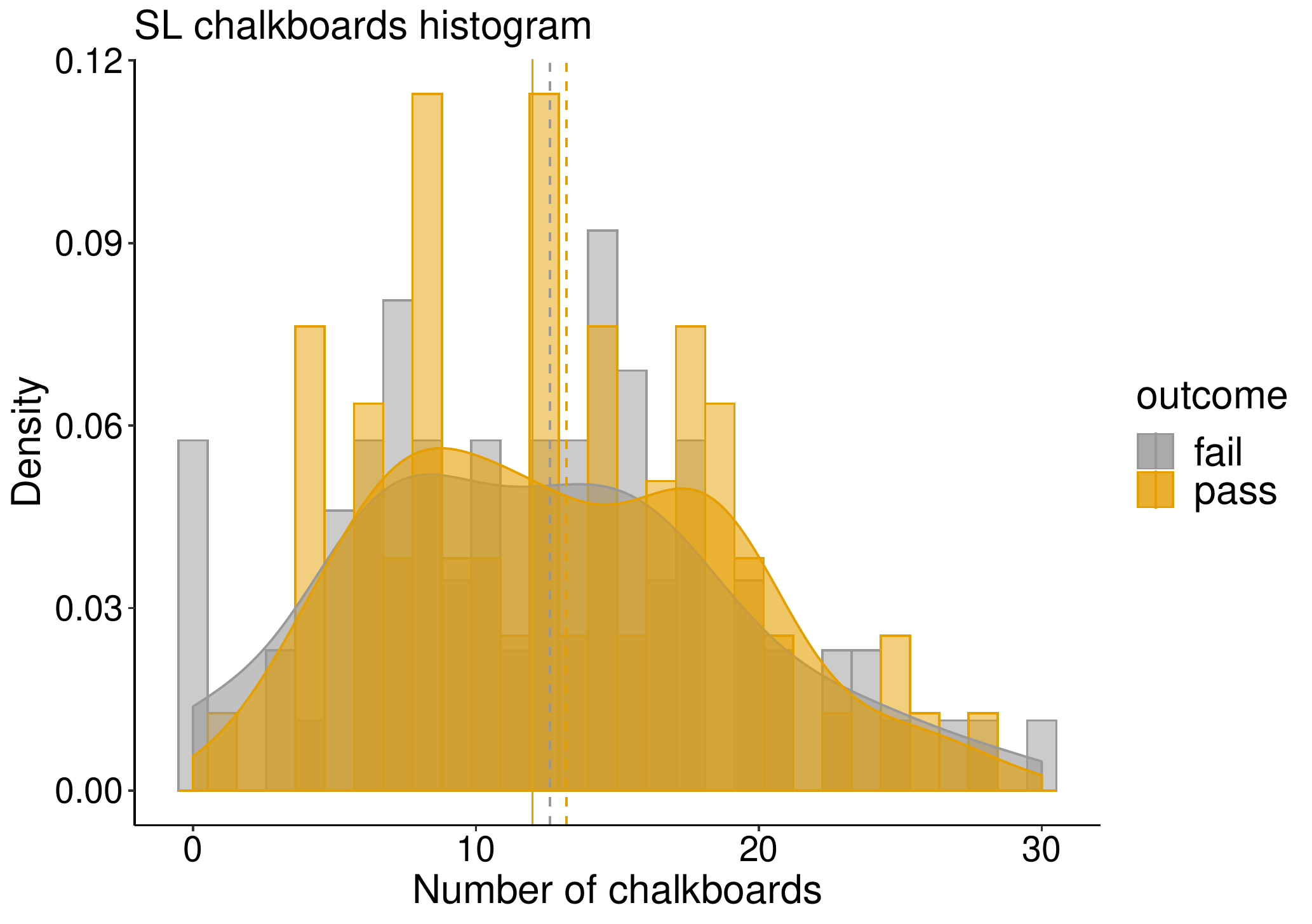}\par 
    \includegraphics[width=.43\textwidth]{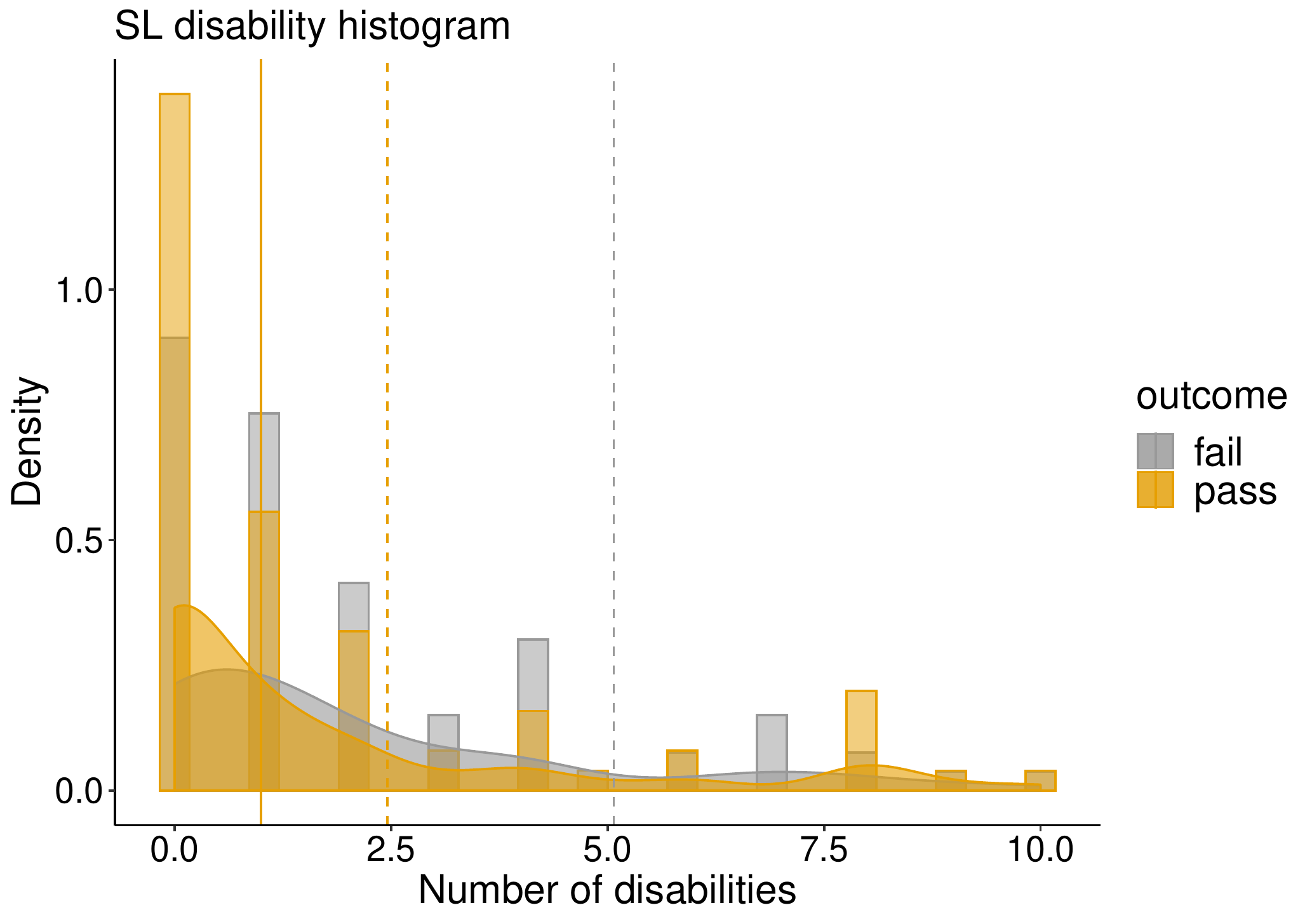}\par 
\end{multicols}
\begin{multicols}{2}
    \includegraphics[width=.43\textwidth]{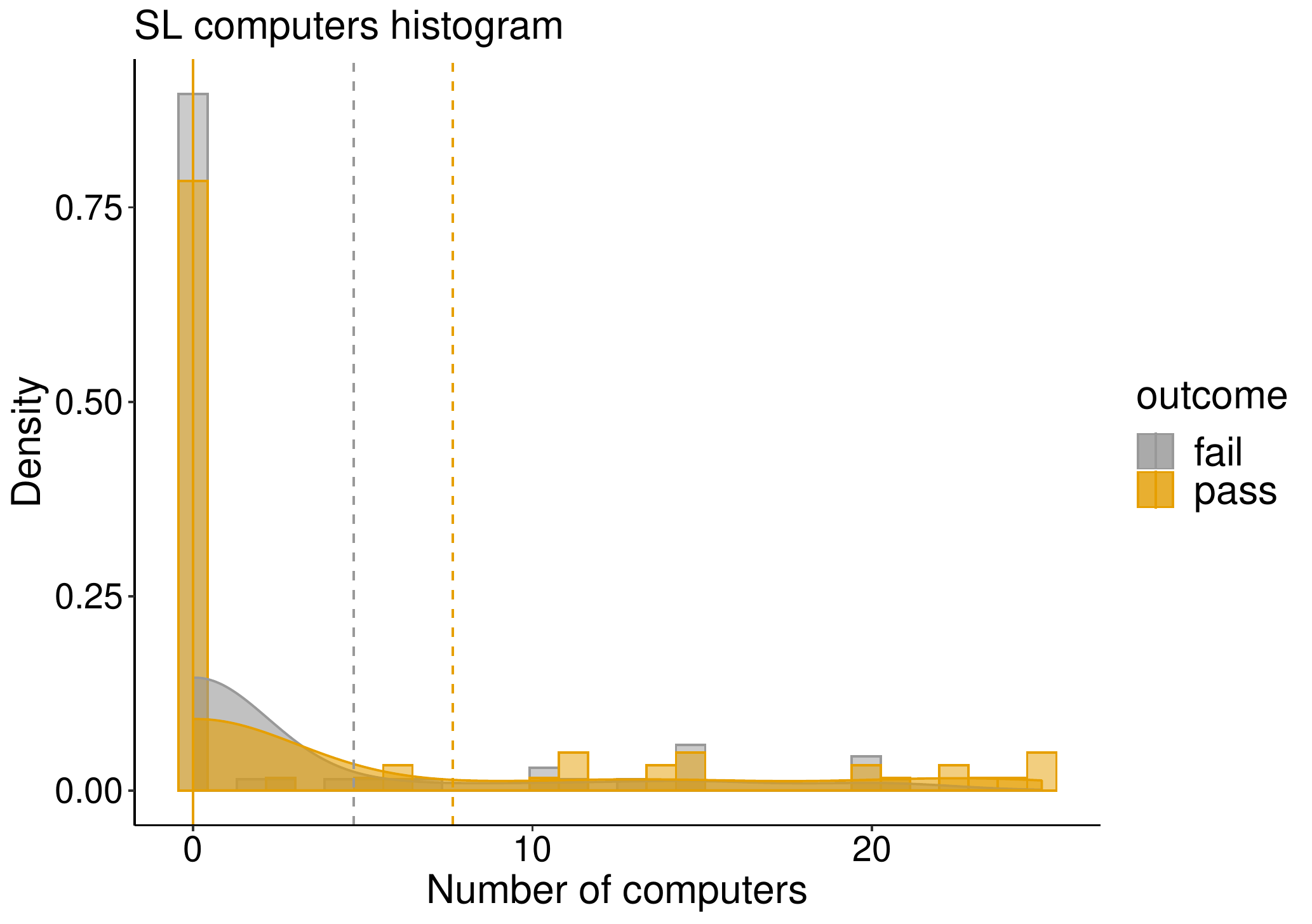}\par
    \includegraphics[width=.43\textwidth]{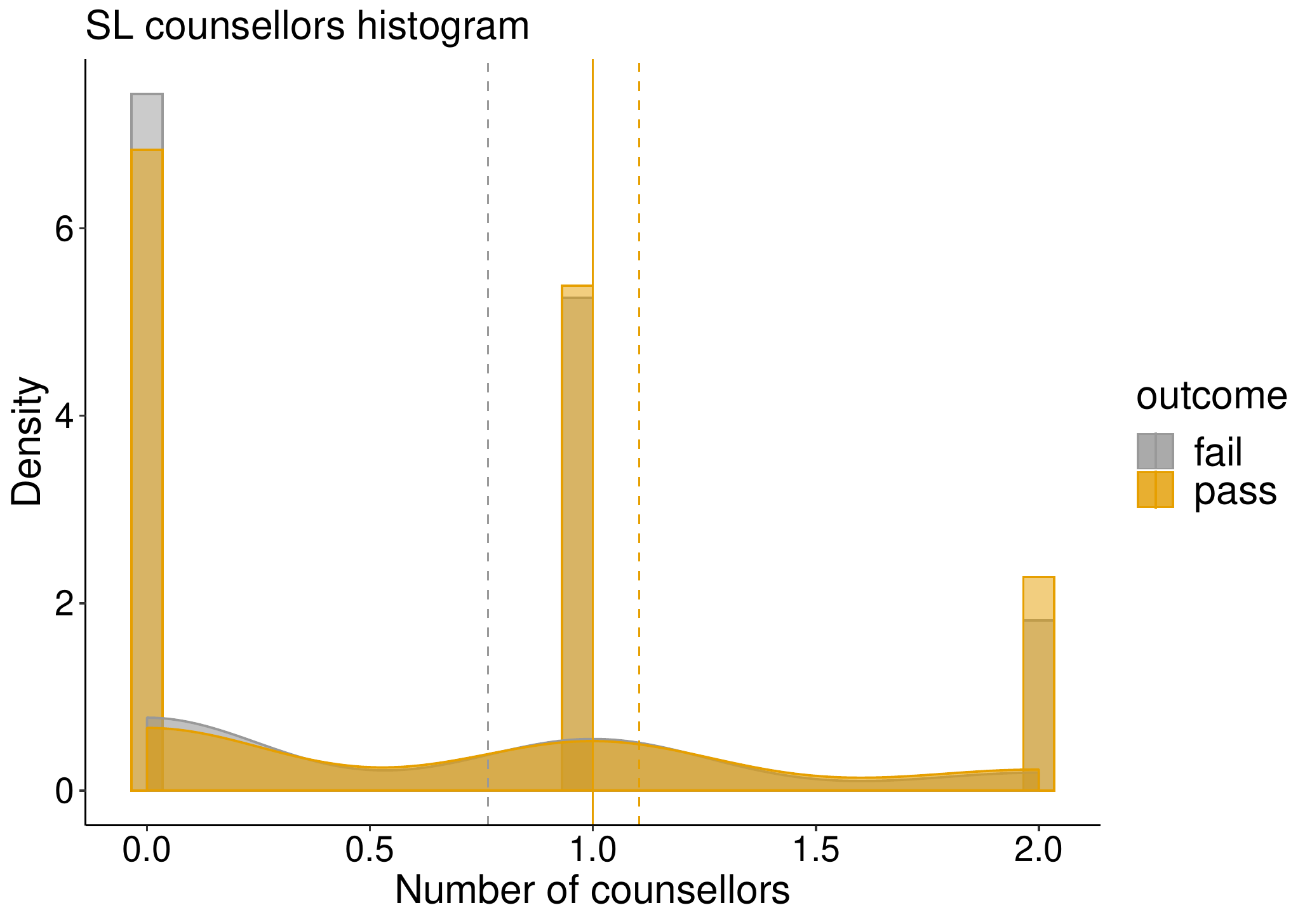}\par
\end{multicols}
\begin{multicols}{2}
    \includegraphics[width=.43\textwidth]{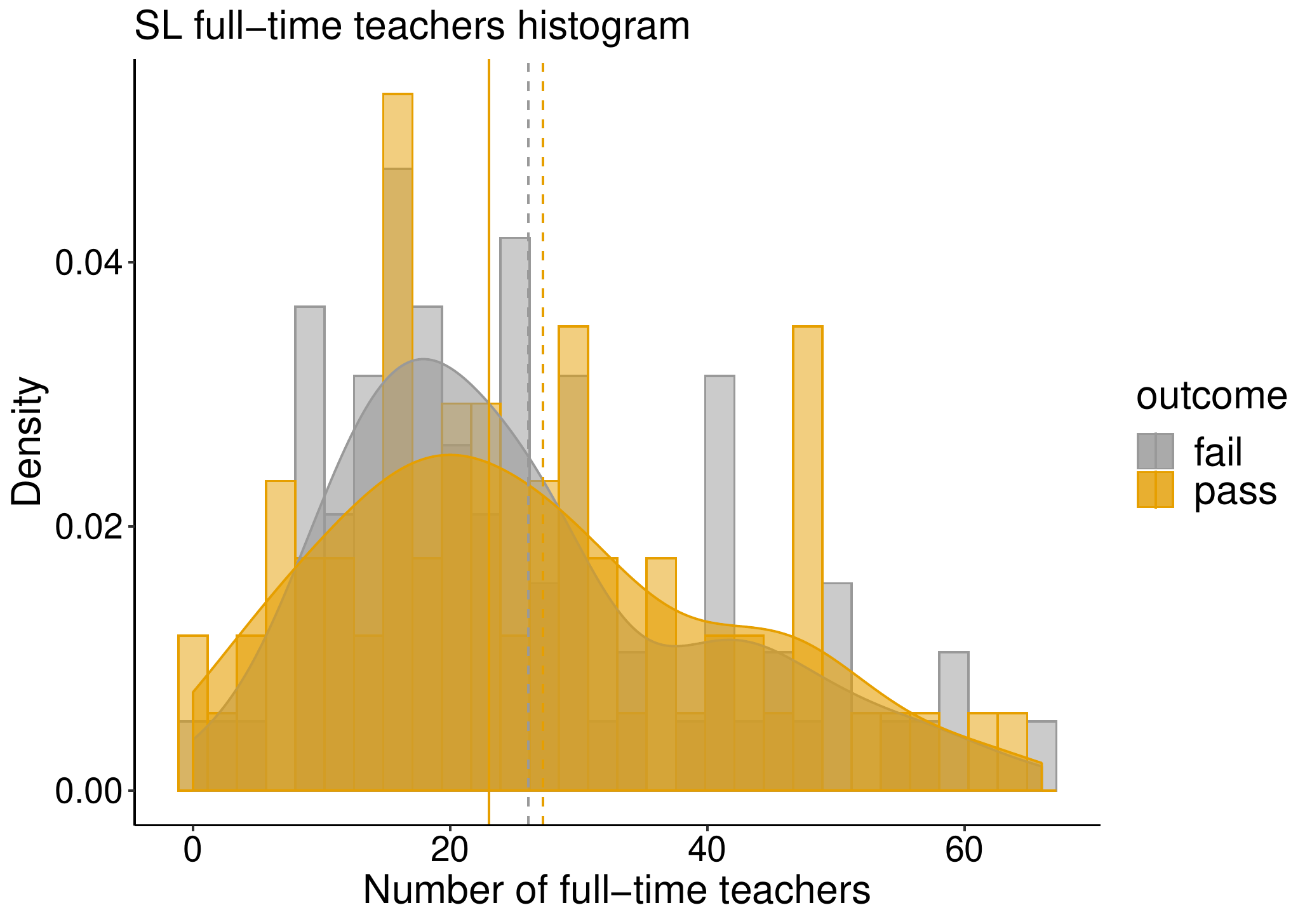}\par
    \includegraphics[width=.43\textwidth]{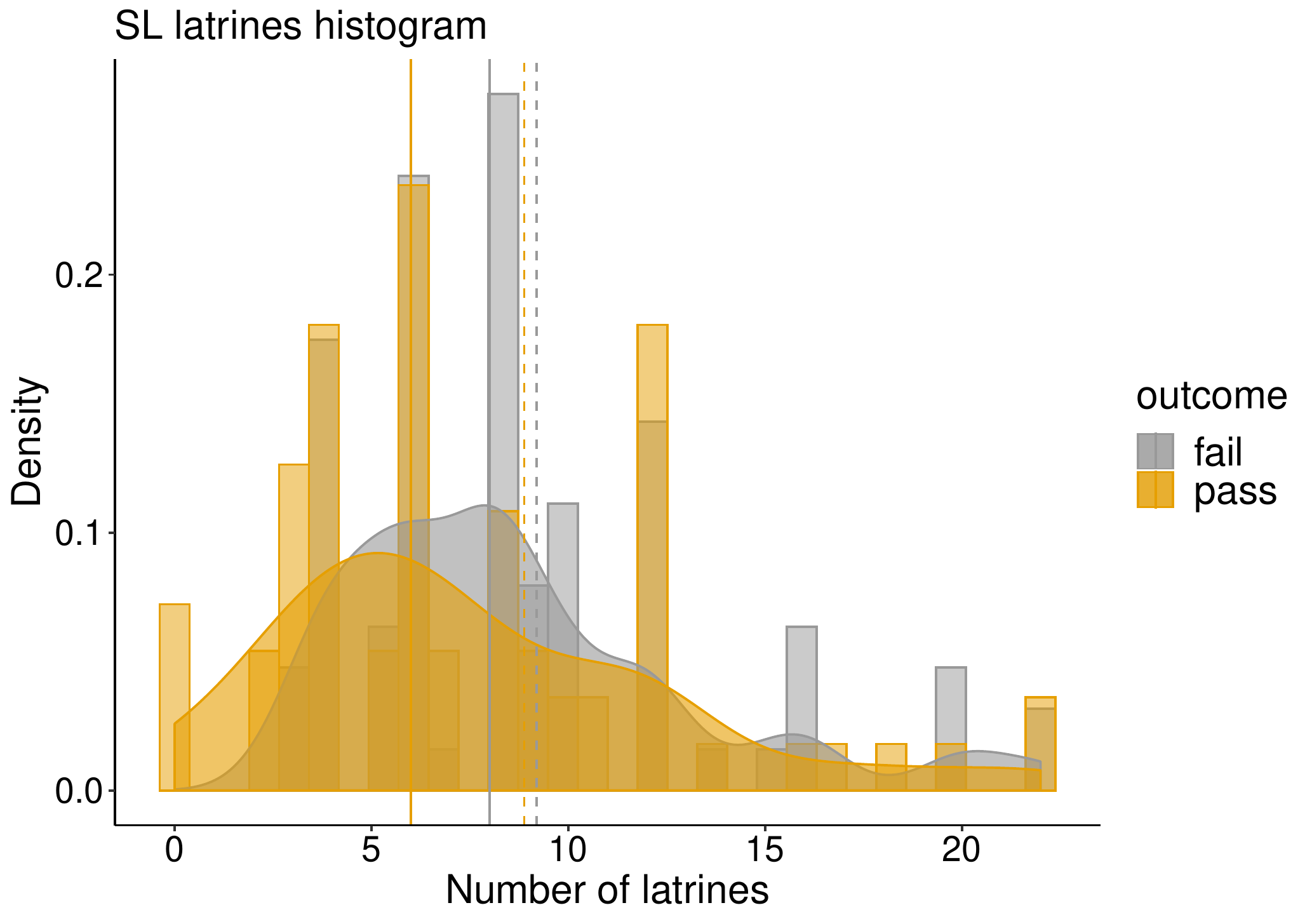}\par
\end{multicols}
\begin{multicols}{2}
    \includegraphics[width=.43\textwidth]{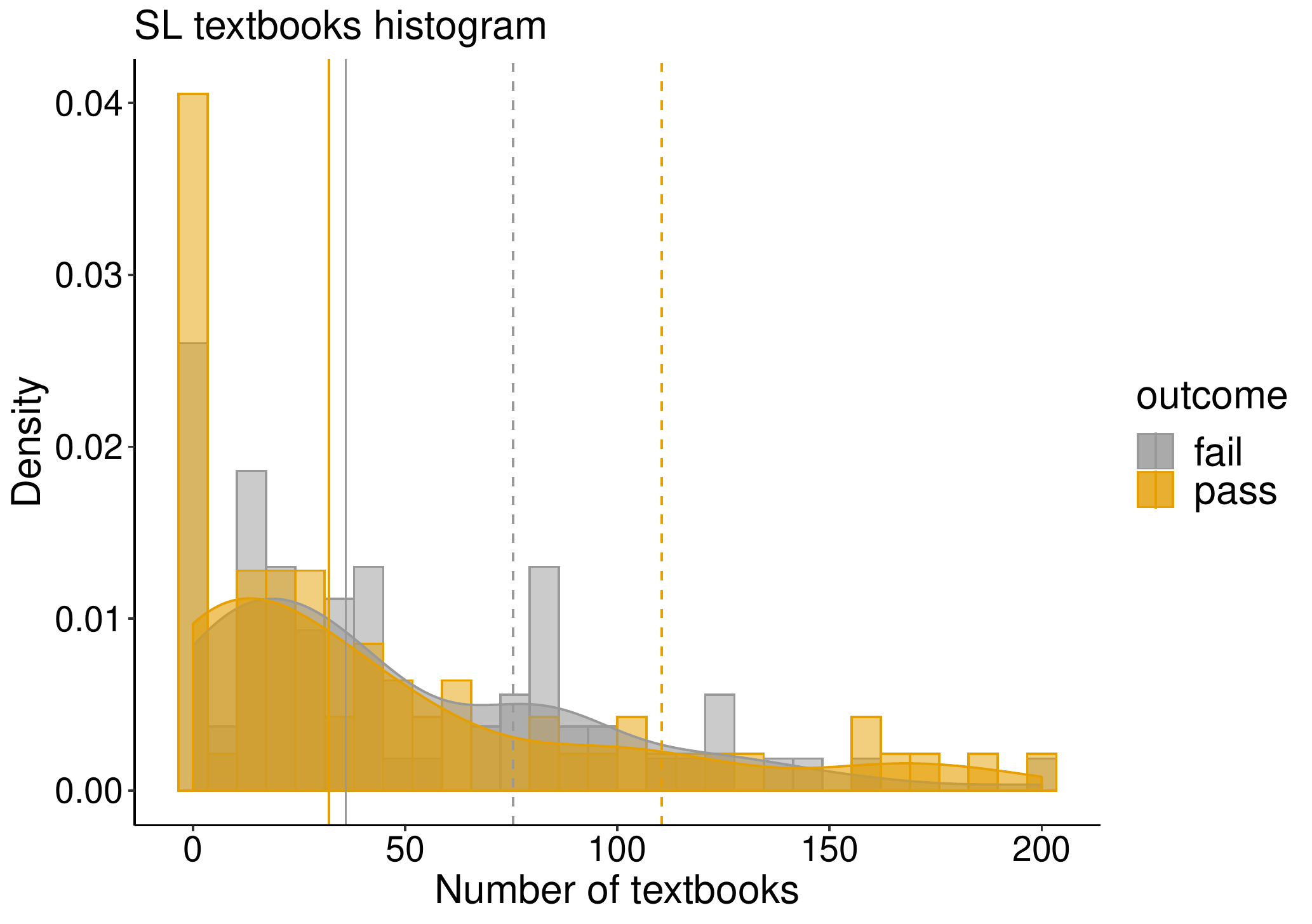}\par
    \includegraphics[width=.43\textwidth]{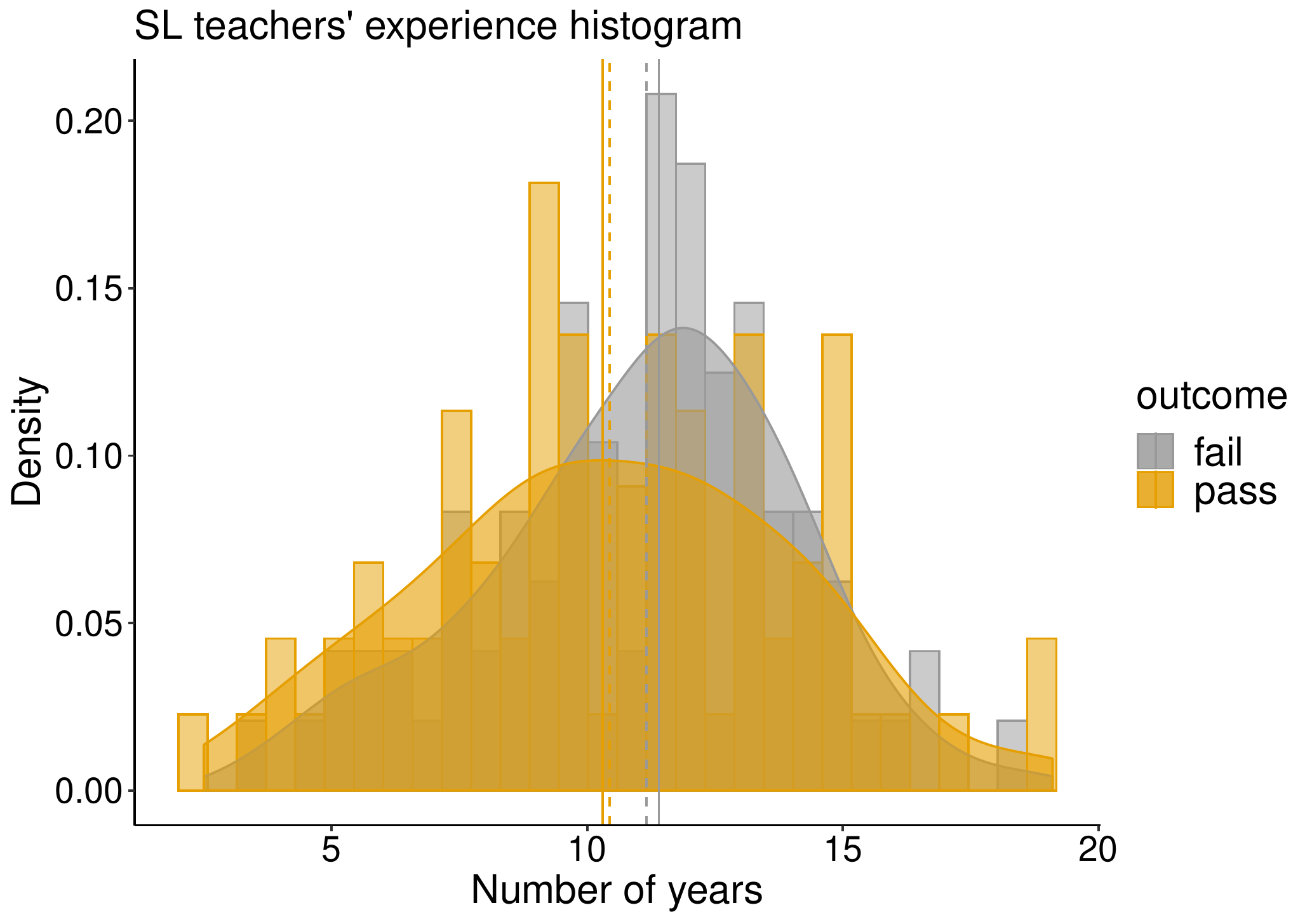}\par
\end{multicols}
\caption{The solid lines represent the median values and average values are represented by dashed lines. It can be noticed that there was no significant differences in the average number school resources in both fail and pass school categories. Considering the median values, all schools had 12 chalkboards, no computers, 6 latrines, and 22 number of full-time teachers.}
\label{fig:sl-numeric-summary}
\end{figure}

\end{document}